 \documentclass[final,1p,times]{elsarticle}

\usepackage{amssymb}
\usepackage{amsmath}

\usepackage{amsmath} 
\usepackage{amsthm}
\usepackage{amsfonts} 
\usepackage{mathrsfs}
\usepackage{lipsum}
\usepackage{subfig}
\usepackage{graphicx} 
\usepackage[export]{adjustbox}
\usepackage{float}
\usepackage{wrapfig}
\usepackage{hyperref} 
\usepackage{lastpage} 
\usepackage{fancyhdr} 
\usepackage{enumitem} 
\usepackage{mathabx} 
\usepackage{booktabs}
\usepackage[skip=0pt]{caption}
\usepackage{comment}
\usepackage{array} 
\newtheorem{theorem}{Theorem}[section]
\usepackage{natbib}
\usepackage{xcolor}

\newlength{\tempheight}
\newlength{\tempwidth}
\newcommand{\rowname}[1]
{\rotatebox{90}{\makebox[\tempheight][c]{\textbf{#1}}}}
\newcommand{\columnname}[1]
{\makebox[\tempwidth][c]{\textbf{#1}}}

\journal{Journal of Theoretical Biology}

\begin{document}
    
\begin{frontmatter}

\title{Stoichiometric ontogenetic development influences population dynamics: Stage-structured model under nutrient co-limitations} 
 
\author[add1]{Tomas Ascoli} 
\author[add2]{Dhruba Pariyar Damay} 
\author[add3]{Jing Li }
\author[add2]{Angela Peace}
\author[add2]{Gregory D. Mayer}
\author[add1]{Rebecca A.  Everett} 

\address[add1]{Haverford College, Haverford, PA, USA}
\address[add2]{Texas Tech University, Lubbock, TX, USA}
\address[add3]{California State University Northridge, Northridge, CA, USA}
 
\begin{abstract}
Ecological processes depend on the flow and balance of essential elements such as carbon (C) and phosphorus (P), and changes in these elements can cause adverse effects to ecosystems. The theory of Ecological Stoichiometry offers a conceptual framework to investigate the impact of elemental imbalances on structured populations while simultaneously considering how ecological structures regulate nutrient cycling and ecosystem processes. While there have been significant advances in the development of stoichiometric food web models, these efforts often consider a homogeneous population and neglect stage-structure. The development of stage-structured population models has significantly contributed to understanding energy flow and population dynamics of ecological systems. However, stage structure models fail to consider food quality in addition to food quantity. We develop a stoichiometric stage-structure producer-grazer model that considers co-limitation of nutrients, and parameterize the model for an algae-\emph{Daphnia} food chain. Our findings emphasize the impact of stoichiometric constraints on structured population dynamics. By incorporating both food quantity and quality into maturation rates, we demonstrate how stage-structured dynamics can influence outcomes in variable environments. Stage-specific parameters, such as juvenile growth and ingestion rates can drive shifts in equilibria, limit cycles, and bifurcation points. These effects are especially significant in high-light environments where nutrient limitations are most pronounced.
\end{abstract}

%


\end{frontmatter}


\section{Introduction}
Stage-structured models are useful for capturing complicated population dynamics \cite{manly1990stage}, however many neglect the role of nutrient availability on stage development. Resource quantity and quality have varying effects on grazer growth at different developmental stages. Early stages, characterized by high growth rates, may be particularly affected by nutrient limitations (e.g., nitrogen N and phosphorus P limitations) \cite{Ecology2004}. Nutrient availability directly impacts grazer growth and ontogeny \cite{Limnol2002, Limnol1998}, and since reproductive tissues are rich in N and P, nutrient-poor food can significantly reduce reproductive output \cite{Freshwater2003}. These stage-specific effects influence population dynamics, underscoring the need for nutrient-dependent, stage-structured modeling approaches.  While there have been significant advances in incorporating nutrient limitations and stoichiometric constraints into population models over the last few decades under the framework of Ecological Stoichiometry \cite{sterner2017ecological}, these efforts often consider homogeneous populations and neglect stage structure. 

The theory of Ecological Stoichiometry consider multiple chemical elements and their ratios across trophic levels to incorporate food quantity and quality into a single framework\cite{sterner2017ecological}. Andersen (1997) \cite{1997andersen} was a pioneer in incorporating stoichiometric effects into mathematical models, modifying the classical Rosenzweig-MacArthur variation of the Lotka-Volterra equations to account for nutrient-limited growth by adjusting both producer growth rates and grazer efficiency. Building on this, Loladze et al. (2000) \cite{2000LKE} developed the LKE model, focusing on a producer-grazer system (algae-Daphnia) and introducing a variable phosphorus-to-carbon (P:C) ratio in producers, termed the cell quota (Q), to reflect food quality. This modeling framework has been expanded to include nutrient excess consequences \cite{2013peace}, multiple nutrients \cite{2002Grover,2003Grover,2004Grover}, explicitly track free nutrients in the environment \cite{2014peace,2008Wang}, determine important trophic transfer efficiencies \cite{2015peace}, and explore dynamic foraging behaviors \cite{2019peace, Oluwagbemisola2024nutrient}. 

Many stoichiometric models utilize threshold minimum functions to account for constraints from multiple elements, following Justus von Liebig’s law of the minimum \cite{sterner2017ecological}.  A smooth alternative is the multiple limitation hypothesis, which proposes that growth can be simultaneously limited by multiple nutrients \cite{MLH}. Liebig’s law assumes that resources are strictly essential, implying that in systems with multiple limiting nutrients, the limiting factor may shift abruptly as nutrient availability changes. In contrast, the multiple limitation hypothesis suggests that nutrients interact, leading to a gradual transition from limitation by one nutrient to another with a region of co-limitation. Liebig's Law may overestimate the growth rate of \textsl{Daphnia magna} when multiple nutrients approach their respective limiting concentrations simultaneously \cite{sperfeld2012multiple}.

The complexities of individual stage structures have largely been neglected in stoichiometric modeling. Current models may fail when fecundity and maturation are crucial in nutrient recycling and ecosystem function. 
To investigate the effects of stoichiometry-dependent organismal stage structures under nutrient constraints, we develop and analyze a stage-structured population model subject to variable nutrients.  Our model incorporates stoichiometric constraints (i.e., nutrient:carbon ratio) into stage-structured population dynamics, which we parameterize for two trophic levels of an aquatic food chain (algae-\textsl{Daphnia}). Here, we consider co-limitation and use smooth functions to incorporate nutrient limitations on growth and maturation rates. The model is described in Section \ref{sec:modelcon}. In Section \ref{sec:modelanalysis} we prove positivity and boundedness of model solutions to ensure biological relevance as well as prove stability of grazer extinction equilibria. Numerical analysis of the model is provided in Section \ref{sec:numanalysis}, including simulations, numerical bifurcation analyses, and parameter sensitivity analysis. Conclusions are discussed in Section \ref{sec:discussion}. 

Our results highlight the roles of stoichiometric constraints on variable producer nutrient content and population dynamics. Incorporating variable food quantity and quality into maturation dynamics showcase the influence that stage structured dynamics can have in variable environments. In our model, stage-specific parameters such as juvenile growth and ingestion rates affect maturation rates and can lead to shifts in equilibria, limit cycles, and bifurcation points. Their affects are particularly pronounced in high light environments where nutrient limitations are most severe.

\section{Model Construction}
\label{sec:modelcon}
Loladze, Kuang, and Elser \cite{2000LKE} formulated a producer-grazer model (LKE model) of the first two trophic levels of an aquatic food chain (algae-\emph{Daphnia}) that incorporated the theory of ecological stoichiometry, given by the following model: 
\begin{subequations}\label{eq:LKE}
\begin{align}
\frac{dx}{dt}=bx\bigg(1-\frac{x}{\min\{K, \frac{P-\theta y}{q}\}}\bigg)-f(x)y, \hspace{100pt}\label{algebra1} \\
\frac{dy}{dt}=\hat{e}\hspace{5pt} \text{min}\bigg\{1, \frac{Q}{\theta}\bigg\}f(x)y-\delta y, \hspace{135pt} \label{algebra2} 
\end{align}
\end{subequations}
where
\begin{equation}
Q=\frac{P-\theta y}{x} \quad \text{and} \quad f(x)=\frac{cx}{a+x},\label{Qeq}
\end{equation}
and where  $x(t)$ and $y(t)$ represent the biomass density of the producer (algae) and grazer (\emph{Daphnia}) respectively, measured in terms of mg C/l. The model assumes that the producer and grazer are composed of two essential elements, C and P and considers their stoichiometric P:C ratios. The model assumes that the producer's P:C ratio $Q$ is variable while the grazer's P:C ratio $\theta$ is constant, and the total phosphorus in the system $P$ is constant. The model further assumes that producers uptake available P immediately and that all the P is either in the producer or grazer populations, allowing the variable $Q$ to be written as given in \eqref{Qeq}. Parameter $b$ is the maximum growth rate of the producer, $K$ is the light-dependent producer carrying capacity in terms of C, $q$ is the producer's minimal P:C ratio required for growth, $\hat{e}<1$ is the grazer's maximum production efficiency, and $\delta$ is the grazer's loss rate. The grazer's ingestion rate $f(x)$ is the Holling type II functional response denoted in \eqref{Qeq}, where $c$ is the maximum ingestion rate and $a$ is the half-saturation constant.

We modify the LKE model \eqref{eq:LKE} to incorporate co-limitation of P and C on producer and grazer growth dynamics using the multiple limitation hypothesis. For the producer's growth, we apply the multiplicative growth rate approach presented by Saito et al. \cite{saito2008some} to replace the minimum function in \eqref{algebra1}. Note that equation \eqref{algebra1} can be written as 
\[ \frac{dx}{dt} = bx\min\left\{1-\frac{x}{K}, 1-\frac{q}{Q}\right\} - f(x)y,\]
which can be written in co-limitation form as 
\[\frac{dx}{dt} = b x \left(1 - \frac{x}{K}\right)\left(1 - \frac{q}{Q}\right) - f(x) y. \label{eq:SmoothLKe_a}\]

\begin{figure}[H]
\subfloat[Leibig producer growth]{\includegraphics[width=.33\textwidth]{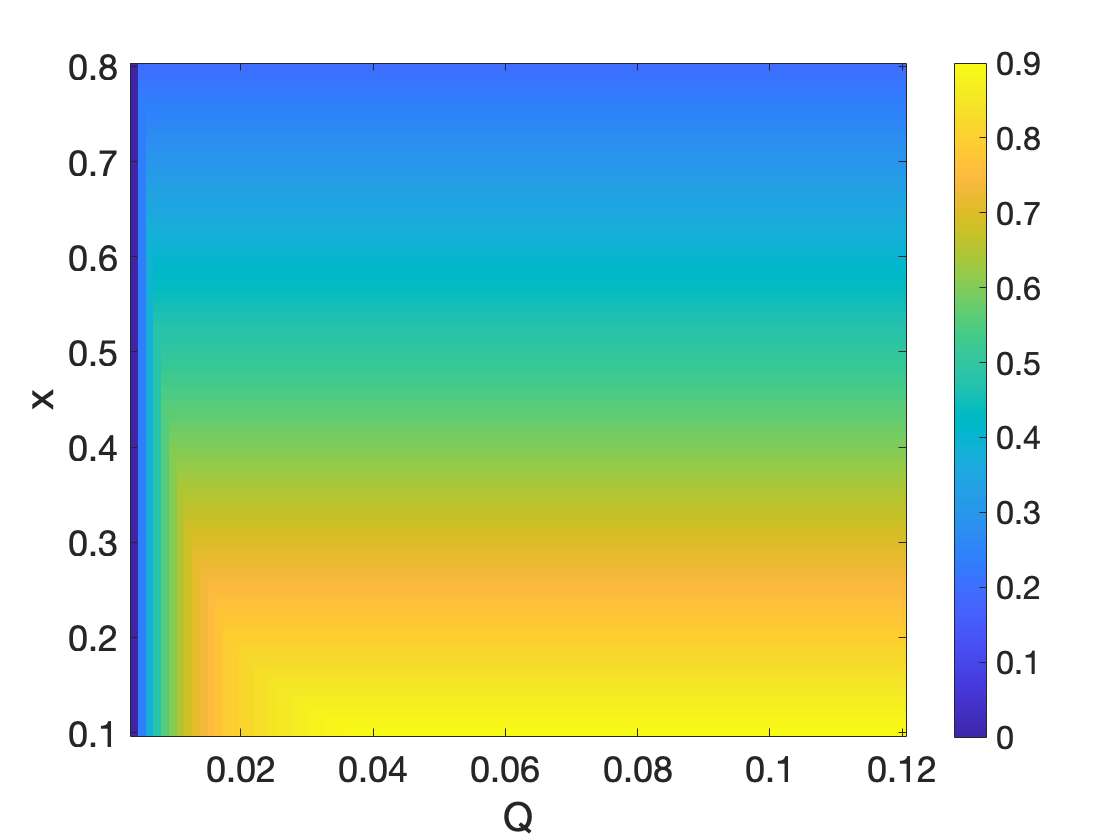}}
 \subfloat[co-limited producer growth]{\includegraphics[width=.33\textwidth]{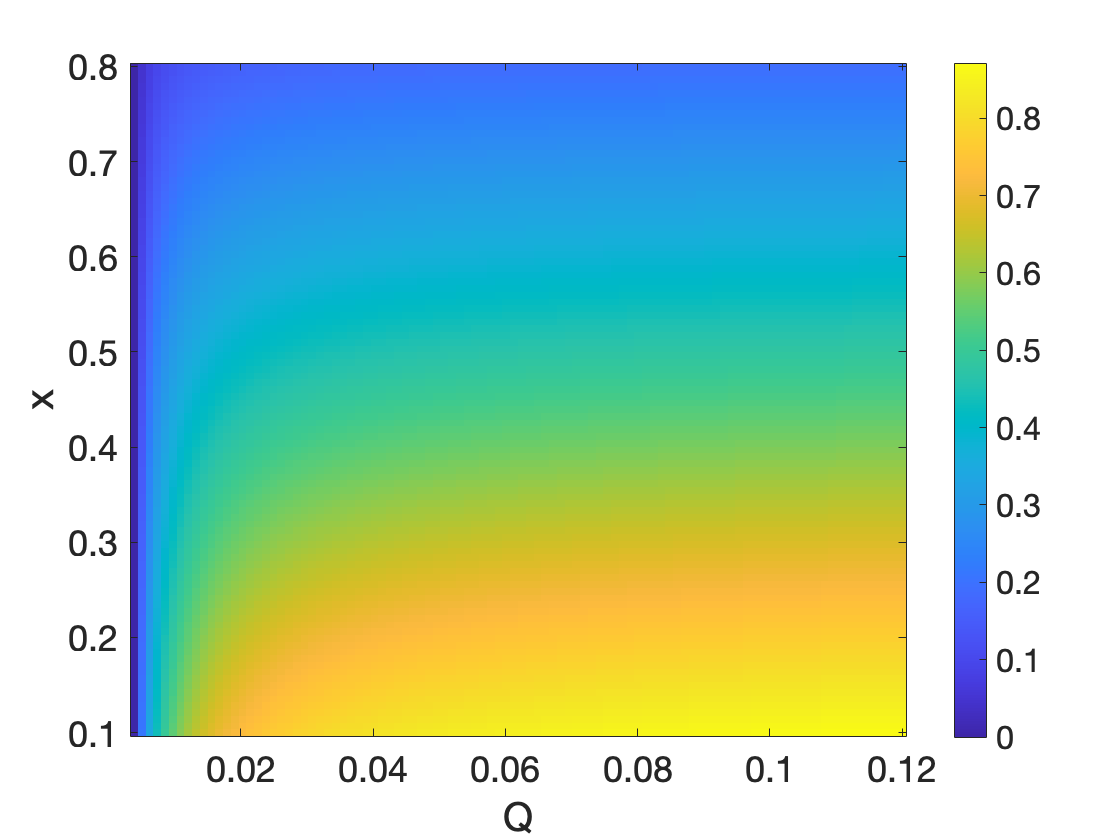}}
\subfloat[grazer conversion efficiencies]{\includegraphics[width=.33\textwidth]{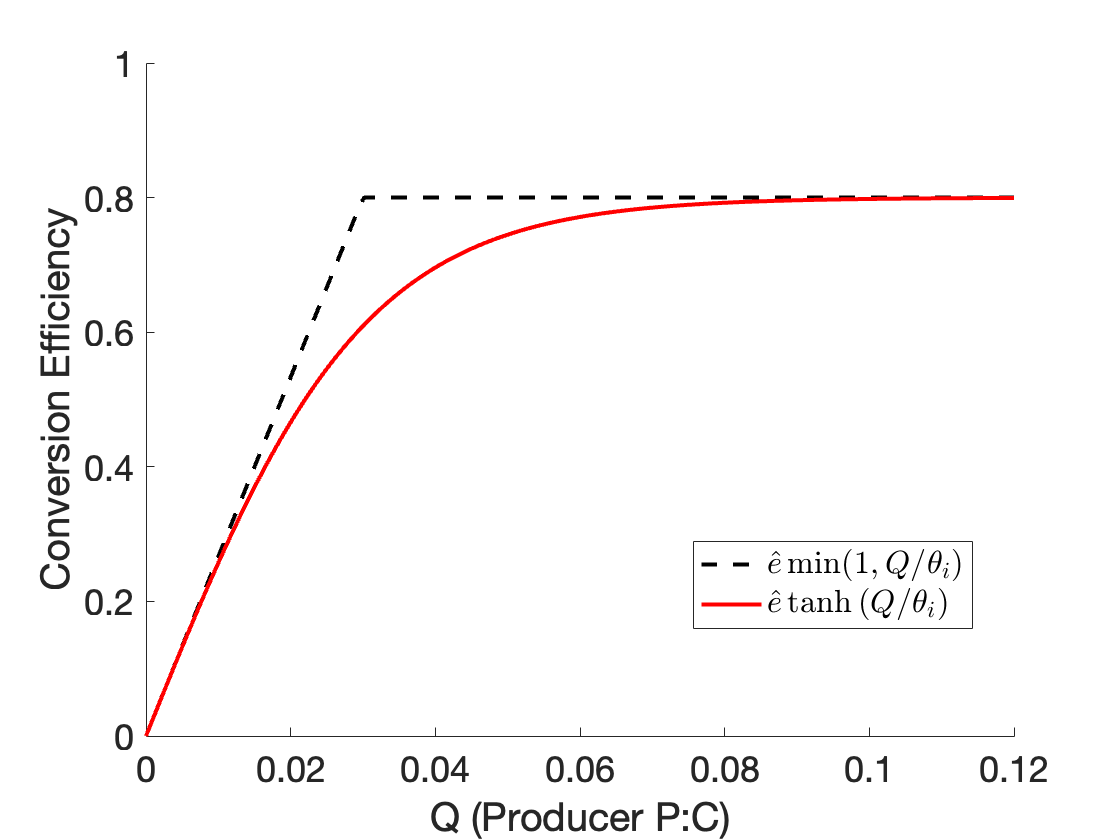}}
 \caption{Visualizations of nonsmooth functional forms using Leibig minimum approach vs smooth co-limitation approach: (a) producer growth with minimum function $\min\left\{1-\frac{x}{K},1-\frac{q}{Q}\right\}$, (b) producer growth using multiplicative co-limitation function $\left(1-\frac{x}{K}\right)\left(1-\frac{q}{Q}\right)$, (c) grazer conversion efficiencies. Here we use $K=2$ mg C/l, $q=0.0038$ mg P/mg C, $\theta=0.03$ mg P/mg C. }
\label{fig:Smooth}
 \end{figure}

We further assume that grazer growth
is co-limited by P and C by replacing the conversion efficiency minimum function in \eqref{algebra2} with the hyperbolic tangent function
\begin{equation}
h(Q) = \tanh{\left(\frac{Q}{\theta}\right)}.
\end{equation}
Note that $h(Q)\leq\frac{Q}{\theta}$.
These changes result in the stoichiometric producer-grazer model with co-limitation given by
\begin{subequations}\label{eq:LKEsmooth}
\begin{align}
    \frac{dx}{dt} &= b x \left(1 - \frac{x}{K}\right)\left(1 - \frac{q}{Q}\right) - f(x) y,\hspace{60pt}\label{eq:SmoothLKe_a} \\
    \frac{dy}{dt} &= \hat{e} h(Q) f(x) y - \delta y. \label{eq:SmoothLKE_b} 
\end{align}
\end{subequations}
Visualizations of these co-limitation functions are given in Figure \ref{fig:Smooth}. 

 We now modify model \eqref{eq:LKEsmooth} to incorporate grazer stage structure by considering the amount of carbon in juvenile $J(t)$ and adult $A(t)$ \emph{Daphnia} populations.  We assume all adult growth goes directly to the reproduction of juveniles, and all juvenile growth goes directly to maturation. These growth dynamics depend directly on ingestion rates and the conversion of food into biomass. 
 The conversion efficiency for reproduction depends on the P:C ratio required to produce juvenilities, $\theta_j$. The conversion efficiency for maturation growth depends on the P:C ratio required by an adult, $\theta_a$. Note that the model does not allow for individual growth outside of maturation or reproduction. Our stoichiometric co-limited stage-structure model is given by 

\begin{subequations}\label{eq:Model}
\begin{align}
    \frac{dx}{dt} &= bx\left(1 - \frac{x}{K}\right)\left(1 - \frac{q}{Q}\right) - f_j(x) J - f_a(x) A,   \label{eq:Model_x} \\
    \frac{dJ}{dt} &= e_a h_j(Q) f_a(x) A - e_j h_a(Q) f_j(x) J - \delta_j J, \label{eq:Model_j} \\
    \frac{dA}{dt} &= e_j h_a(Q) f_j(x) J - \delta_a A, \label{eq:Model_a}
\end{align}
\end{subequations}
where $Q = \frac{P - \theta_j J - \theta_a A}{x}$, $f_i(x)=\frac{c_ix}{a_i+x}$, and  $h_i(Q) = \tanh{\left(\frac{Q}{\theta_i}\right)}$ for $i=j$, $a$ for juveniles and adults respectively.  We now have stage-specific parameters $c_i$, $a_i$, $e_i$, $\theta_i$, and $\delta_i$, for $i=j$, $a$. A list of the parameters are given in Table \ref{tab:params}.

\section{Model Analysis}
\label{sec:modelanalysis}
Here, we conduct basic analyses of our stoichiometric co-limited stage-structure model \eqref{eq:Model} verifying solutions remain positive and bounded in biologically realistic ranges with the following theorem. 

\begin{theorem}
\label{thm:smoothstagebounded} Solution to the system \eqref{eq:Model} with initial conditions in the set
$$\Omega=\bigg\{(x,J,A):0\leq x \leq \min\bigg\{K, \frac{P}{q}\bigg\},0\leq J, 0\leq A, qx+\theta_j J+ \theta_a A < P 
 \bigg\}$$
will remain there for all forward time.
\end{theorem}
\medskip
\begin{proof}
Let $S(t)=(x(t), J(t),A(t))$ be a solution of system \eqref{eq:Model}
 with $S(0) \in \Omega$. Assume that $S(t_1)$ touches or crosses a boundary of $\Omega$ for the first time for some $t_1 >0$. We prove this in the following cases.
 \begin{enumerate}
 \item[] \textbf{Case 1:} If $J(t_1)=0$ for $t_1 >0$. Since $A(t_1) \geq 0$, then $\frac{dJ}{dt}|_{t=t_1} \geq 0$, and hence $J(t)\geq 0$  for all $t>t_1>0$ or we can say $J(t)\geq 0$ for all $t>0$.
 \item[] \textbf{Case 2:} If $A(t_1)=0$ for $t_1 >0$. Since $J(t_1) \geq 0$, then $\frac{dA}{dt}|_{t=t_1} \geq 0$, and hence $A(t)\geq 0$  for all $t>t_1>0$ or we can say $A(t)\geq 0$ for all $t>0$.
     \item[] \textbf{Case 3:} If $x(t_1)=0 $ for $t_1 >0$. Then $\frac{dx}{dt}|_{t=t_1} = 0$, and hence $x(t)= 0$  for all $t>t_1>0$ or we can say $x(t)\geq 0$ for all $t>0$.
    \item[] \textbf{Case 4:} If $x(t_1)=\min\{K, \frac{P}{q}\}$ for $t_1>0$. Then, for every $t \in [0, t_1]$,
     \begin{eqnarray*}
     x'&=&bx\left(1 - \frac{x}{K}\right)\left(1 - \frac{q}{Q(t)}\right)-f_j(x)J-f_a(x)A \\
     &\leq& bx\left(1 - \frac{x}{K}\right)\left(1 - \frac{q}{Q(t)}\right) \\
      &=& bx\left(1 - \frac{x}{K}\right)\left(1 - \frac{x}{\frac{P-\theta_j J-\theta_a A}{q}}\right) \\
    &\leq& bx\left(1 - \frac{x}{K}\right)\left(1 - \frac{x}{\frac{P}{q}}\right). \\
    \end{eqnarray*}
Therefore $x'\leq bx\left(1-\frac{x}{K}\right)$ and $x'\leq bx\left(1-\frac{x}{\frac{P}{q}}\right)$. 
   By the standard comparison argument, $x(t_1)\leq \min\{K, \frac{P}{q}\}$. Thus $S(t_1)$ can not cross this boundary. 
\item[] \textbf{Case 5:} Suppose the inequality $qx+\theta_j J+\theta_a A < P$ is not true for the first time for $t_1>0$\\ and let 
\begin{equation}\label{eq:C1}
qx(t_1)+\theta_jJ(t_1)+\theta_aA(t_1)=P, \hspace{5pt} \text{for}\hspace{5pt}  t_1>0.
\end{equation}
 Since for $t\in [0,t_1)$, $qx(t)+\theta_j J(t)+\theta_a A(t) < P$, implies that 
 \begin{equation}\label{eq:C2}
 qx'(t_1)+\theta_j J'(t_1)+\theta_a A'(t_1) \geq 0.
 \end{equation}
 From $\eqref{eq:C1}$, it follows that $$x(t_1)=\frac{P-\theta_j J(t_1)-\theta_a A(t_1)}{q} \hspace{5pt} \text{and} \hspace{5pt} q=\frac{P-\theta_j J(t_1)-\theta_a A(t_1)}{x(t_1)}=Q(t_1). $$
 From $\eqref{eq:Model_x}$, to obtain bounds for $x'(t_1):$
 \begin{eqnarray}\label{eq:C3}
    x'(t_1)&=&bx(t_1)\left(1 - \frac{x(t_1)}{K}\right)\left(1 - \frac{q}{Q(t_1)}\right) - f_j(x(t_1)) J(t_1) - f_a(x(t_1)) A(t_1) \nonumber\\
 &=& -f_j(x(t_1)) J(t_1) - f_a(x(t_1)) A(t_1).
  \end{eqnarray}
From $\eqref{eq:Model_j}$, to obtain bounds for $J'(t_1):$
\begin{eqnarray}\label{eq:C4}
J'(t_1)&=&e_a h_j(Q(t_1)) f_a(x(t_1)) A(t_1) - e_j h_a(Q(t_1)) f_j(x(t_1)) J(t_1) - \delta_j J(t_1) \nonumber\\
&\leq& e_a h_j(Q(t_1)) f_a(x(t_1)) A(t_1) 
\nonumber\\
&\leq& e_a\frac{q}{\theta_j}f_a(x(t_1))A(t_1).
\end{eqnarray}
From $\eqref{eq:Model_a}$, to obtain bounds for $A'(t_1):$
\begin{equation}\label{eq:C5}
A'(t_1)=e_j h_a(Q(t_1)) f_j(x(t_1)) J(t_1) - \delta_a A(t_1)\leq e_j\frac{q}{\theta_a}f_j(x(t_1))J(t_1).
\end{equation}
Using $\eqref{eq:C3}, \eqref{eq:C4}, \eqref{eq:C5}$ and the fact that $e_i<1$, for $i=j, a$, we obtain the following:
\begin{align*}
    q&x'(t_1)+\theta_j J'(t_1)+\theta_a A'(t_1) \\
    &\leq q\left(-f_j(x(t_1)) J(t_1) - f_a(x(t_1)) A(t_1)\right)
    +\theta_j e_a\frac{q}{\theta_j}f_a(x(t_1)) A(t_1)+\theta_a e_j\frac{q}{\theta_a}f_j(x(t_1))J(t_1) \nonumber \\
    &\leq q(e_j-1)f_j(x(t_1))J(t_1)+q(e_a-1)f_a(x(t_1))A(t_1) <0
\end{align*}
This contradicts $\eqref{eq:C2}$ and completes the proof.
   \end{enumerate}
   \end{proof}
   
 Model \eqref{eq:Model} has two equilibria where the grazer population dies out, $E_1=(K,0,0)$ and $E_2=\left(\frac{P}{q},0,0\right)$. The following theorem presents results on the local stability of these equilibria. 

\begin{theorem}
The grazer extinction equilibria $E_1$ and $E_2$ of the stoichiometric co-limited stage-structure model \eqref{eq:Model} have the following properties:
\begin{enumerate}
    \item[(a)] $E_1=(K,0,0)$ is locally asymptotically stable if 
$$K<\frac{P}{q} \hspace{5pt} \text{and} \hspace{5pt} e_jh_a\left(\frac{P}{K}\right) f_j(K) \bigg(e_a h_j\left(\frac{P}{K}\right) f_a(K) -\delta_a\bigg)< \delta_j \delta_a,$$ and $E_1$ is unstable if $$K>\frac{P}{q} \hspace{5pt} \text{or} \hspace{5pt} e_jh_a\left(\frac{P}{K}\right) f_j(K) \bigg(e_a h_j\left(\frac{P}{K}\right) f_a(K) -\delta_a\bigg)>\delta_j \delta_a.$$
    \item[(b)] $E_2=(\frac{P}{q},0,0)$ is locally asymptotically stable if 
$$K>\frac{P}{q} \hspace{5pt} \text{and} \hspace{5pt} e_j h_a\left(q\right) f_j\bigg(\frac{P}{q}\bigg) \bigg(e_a h_j\left(q\right)f_a\bigg(\frac{P}{q}\bigg) -\delta_a\bigg)< \delta_j \delta_a,$$ and $E_2$ is unstable if $$K<\frac{P}{q} \hspace{5pt} \text{or} \hspace{5pt} e_j h_a\left(q\right) f_j\bigg(\frac{P}{q}\bigg) \bigg(e_a h_j\left(q\right)f_a\bigg(\frac{P}{q}\bigg) -\delta_a\bigg)> \delta_j \delta_a.$$
\end{enumerate}
\end{theorem}
\medskip
\begin{proof}
We first rewrite System \eqref{eq:Model} as
\begin{subequations}\label{eq:system8}
\begin{align}
x' &= b x G(x) N(x,J,A) - f_j (x) J - f_a (x) A  \label{eq:1a}\\
J' &= e_a R(x,J,A) f_a (x) A - e_j M(x,J,A) f_j (x) J - \delta_j J\label{eq:8b}\\
 A' &= e_j M(x,J,A) f_j (x) J - \delta_a A, \label{eq:8c}
\end{align}
\end{subequations}
where $G(x) = 1 - \frac{x}{K}$, $N(x,J,A) = 1 - \frac{q x}{P - \theta_j J - \theta_a A}$,  $R(x,J,A) = h_j(Q)$, and $M(x,J,A) = h_a(Q)$.\\
The Jacobian of System \eqref{eq:system8} is:
\begin{equation}\label{eq:Jac}
\mathcal{J} = 
\begin{bmatrix}
    \mathcal{J}_{11} & \mathcal{J}_{12}  &\mathcal{J}_{13} \\
     \mathcal{J}_{21} & \mathcal{J}_{22}  &\mathcal{J}_{23} \\
      \mathcal{J}_{31} & \mathcal{J}_{32}  &\mathcal{J}_{33} 
\end{bmatrix}
\end{equation}
where $\mathcal{J}_{11}= b G N + b x G' N + b x G N_x - f'_j J - f'_a A, \mathcal{J}_{12}=b x G N_J - f_j, \mathcal{J}_{13}=b x G N_A - f_a, \mathcal{J}_{21}=e_a A \left( R_x f_a + R f'_a\right) - e_j J \left(M_x f_j + M f'_j\right), \mathcal{J}_{22}=e_a R_J f_a A - e_j f_j \left(M_J J + M\right) - \delta_j, \mathcal{J}_{23}=e_a f_a \left(R_A A + R \right) - e_j M_A f_j J, \mathcal{J}_{31}=e_j J \left(M_x f_j + M f'_j\right), \mathcal{J}_{32}= e_j f_j \left(M_J J + M\right), \mathcal{J}_{33}=e_j M_A f_j J - \delta_a$. 



\begin{itemize}
\item[] \textbf{To prove (a):} We evaluate the Jacobian in \eqref{eq:Jac} at $E_1$, 
\begin{equation*}
\mathcal{J}\bigg|_{E_1} = 
\begin{bmatrix}
    -b \left(1 - \frac{K q}{P}\right) && -f_j(K) && -f_a(K) \\
    0 && -e_j f_j(K) M(K,0,0) - \delta_j && e_a f_a R(K,0,0) \\
    0 && e_j f_j(K) M(K,0,0) && -\delta_a
\end{bmatrix}.
\end{equation*}
Let $\alpha_1 = e_j f_j(K) M(K,0,0)>0$ and $\alpha_2 = e_a f_a(K) R(K,0,0)>0$. To calculate its eigenvalues, we obtain 
the corresponding characteristic equation 
$$\big\{-b \left(1 - \frac{K q}{P}\right)-\lambda\big\}\bigg\{(-\alpha_1-\delta_j-\lambda)(-\delta_a-\lambda)-\alpha_1 \alpha_2 \bigg\}=0,$$
or equivalently,
$$\bigg\{b \left(1 - \frac{K}{\frac{P}{q}}\right)+\lambda\bigg\}\bigg(\lambda^2 +(\alpha_1+\delta_j+\delta_a)\lambda +\alpha_1(\delta_a-\alpha_2)+\delta_j \delta_a\bigg)=0.$$
Then either $b \left(1 - \frac{K}{\frac{P}{q}}\right)+\lambda=0,$ which implies  $\lambda=-b \left(1 - \frac{K}{\frac{P}{q}}\right)<0$ if  $K<\frac{P}{q}$,\\
or $\lambda^2 +(\alpha_1+\delta_j+\delta_a)\lambda +\alpha_1(\delta_a-\alpha_2)+\delta_j \delta_a=0$.\\\\
Let $$\mathcal{P}(\lambda)=\lambda^2 +(\alpha_1+\delta_j+\delta_a)\lambda +\alpha_1(\delta_a-\alpha_2)+\delta_j \delta_a.$$
Since $\alpha_1=e_j f_j(K) M(K,0,0)>0$, $\alpha_2=e_a f_a(K) R(K,0,0)>0$, and $\delta_j, \delta_a>0$, then $\alpha_1+\delta_j+\delta_a>0$. If $\alpha_1(\delta_a-\alpha_2)+\delta_j \delta_a>0$, then by the Routh-Hurwitz Criteria, all of the roots of polynomial $\mathcal{P}(\lambda)$ are negative or have negative real parts. Thus $E_1$ is locally asymptotically stable if $$\alpha_1(\alpha_2-\delta_a)<\delta_j \delta_a,$$ 
which simplifies to
$$ e_jh_a\left(\frac{P}{K}\right) f_j(K) \bigg(e_a h_j\left(\frac{P}{K}\right) f_a(K) -\delta_a\bigg)< \delta_j \delta_a. $$
Thus, $E_1$ is locally asymptotically stable if $$K<\frac{P}{q} \hspace{5pt} \text{and} \hspace{5pt} e_jh_a\left(\frac{P}{K}\right) f_j(K) \bigg(e_a h_j\left(\frac{P}{K}\right) f_a(K) -\delta_a\bigg)< \delta_j \delta_a.$$
Also note that $E_1$ is unstable if $$K>\frac{P}{q} \hspace{5pt} \text{or} \hspace{5pt} e_jh_a\left(\frac{P}{K}\right) f_j(K) \bigg(e_a h_j\left(\frac{P}{K}\right) f_a(K) -\delta_a\bigg)>\delta_j \delta_a.$$
\item[] \textbf{To prove (b):}  We evaluate the Jacobian in \eqref{eq:Jac} at $E_2$, 
$$
\mathcal{J}\bigg|_{E_2} =
\begin{bmatrix}
    -b \left(1 - \frac{P}{Kq}\right) && -f_j(\frac{P}{q}) - \mu \theta_j && -f_a(\frac{P}{q}) - \mu \theta_a \\
    0 && -e_j f_j(\frac{P}{q}) M(\frac{P}{q},0,0) - \delta_j && e_a f_a(\frac{P}{q}) R(\frac{P}{q},0,0) \\
    0 && e_a f_a(\frac{P}{q}) R(\frac{P}{q},0,0) && -\delta_a
\end{bmatrix},
$$
where  $\mu = \frac{b}{q} \left(1 - \frac{P}{K q}\right)$. Let $\bar{\alpha}_1 = e_j f_j(\frac{P}{q}) M(\frac{P}{q},0,0)>0$ and $\bar{\alpha}_2 = e_a f_a(\frac{P}{q}) R(\frac{P}{q},0,0)>0$. To calculate its eigenvalues, we obtain
the corresponding characteristic equation 
$$\big\{-b \left(1 - \frac{\frac{P}{q}}{K}\right)-\lambda\big\}\bigg\{(-\bar\alpha_1-\delta_j-\lambda)(-\delta_a-\lambda)-\bar\alpha_1 \bar\alpha_2 \bigg\}=0,$$ or equivalently, 
$$\bigg\{b \left(1 - \frac{\frac{P}{q}}{K}\right)+\lambda\bigg\}\bigg(\lambda^2 +(\bar\alpha_1+\delta_j+\delta_a)\lambda +\bar\alpha_1(\delta_a-\bar\alpha_2)+\delta_j \delta_a\bigg)=0.$$
Then either $b \left(1 - \frac{\frac{P}{q}}{K}\right)+\lambda=0$, which implies $\lambda=-b \left(1 - \frac{\frac{P}{q}}{K}\right)<0$ if  $K>\frac{P}{q}$,\\
or $\lambda^2 +(\bar\alpha_1+\delta_j+\delta_a)\lambda +\bar\alpha_1(\delta_a-\bar\alpha_2)+\delta_j \delta_a=0$.\\\\
Let $$\bar{\mathcal{P}}(\lambda)=\lambda^2 +(\bar\alpha_1+\delta_j+\delta_a)\lambda +\bar\alpha_1(\delta_a-\bar\alpha_2)+\delta_j \delta_a.$$
Since $\bar\alpha_1=e_j f_j(\frac{P}{q}) M(\frac{P}{q},0,0)>0$, $\bar{\alpha}_2=e_a f_a(\frac{P}{q}) R(\frac{P}{q},0,0)>0$, and $\delta_j, \delta_a>0$, then $\bar\alpha_1+\delta_j+\delta_a>0$. 
If $\bar\alpha_1(\delta_a-\bar\alpha_2)+\delta_j \delta_a>0$, then by the Routh-Hurwitz Criteria, all of the roots of polynomial $\bar{\mathcal{P}}(\lambda)$ are negative or have negative real parts. Thus $E_2$ is locally asymptotically stable if $$\bar\alpha_1(\bar\alpha_2-\delta_a)<\delta_j \delta_a,$$ 
which simplifies to
$$ e_j h_a\left(q\right) f_j\bigg(\frac{P}{q}\bigg) \bigg(e_a h_j\left(q\right)f_a\bigg(\frac{P}{q}\bigg) -\delta_a\bigg)< \delta_j \delta_a. $$
Thus, $E_2$ is locally asymptotically stable if $$K>\frac{P}{q} \hspace{5pt} \text{and} \hspace{5pt} e_j h_a\left(q\right) f_j\bigg(\frac{P}{q}\bigg) \bigg(e_a h_j\left(q\right)f_a\bigg(\frac{P}{q}\bigg) -\delta_a\bigg)< \delta_j \delta_a.$$
Also note that $E_2$ is unstable if $$K<\frac{P}{q} \hspace{5pt} \text{or} \hspace{5pt} e_j h_a\left(q\right) f_j\bigg(\frac{P}{q}\bigg) \bigg(e_a h_j\left(q\right)f_a\bigg(\frac{P}{q}\bigg) -\delta_a\bigg)> \delta_j \delta_a.$$
\end{itemize}
\end{proof}

\section{Numerical Analysis}
\label{sec:numanalysis}
This section describes the results of numerical experiments, numerical bifurcation analyses, and parameter sensitivity analyses of the stoichiometric co-limited stage-structure model \eqref{eq:Model}. 
Baseline parameter values and ranges are shown in Table \ref{tab:params}. Many of the parameter values were those used in the LKE model \eqref{eq:LKE} \cite{2000LKE}. The juvenile maximum production efficiency ($e_j$) and juvenile constant P:C ratio ($\theta_j$) baseline values were assumed, and then we explored varying $e_j$ and $\theta_j$ values in the numerical simulations and sensitivity analysis. The loss rates ($\delta_j, \delta_a$) include both natural death rate and outside factors such as predation. The baseline values were assumed and then we considered a range of loss rate values in the sensitivity analysis. The juvenile ingestion half-saturation constant ($a_j$) was assumed to be the same as the adult ingestion half-saturation constant ($a_a$). Lastly, following \cite{2008McCauley}, we assumed the juvenile maximum ingestion rate ($c_j$) is smaller than the adult maximum ingestion rate ($c_a$), and explored varying $c_j$ values in the numerical simulations and sensitivity analysis.

\begin{table}[h]
\centering
\begin{tabular}{|c|p{4cm}|l|l|c|}
    \hline
   \textbf{Parameter}  & \textbf{Description} & \textbf{Baseline} & \textbf{Range} & \textbf{Source}\\
   \hline
    $ P$ &  Total phosphorus in system  & 0.025 mg P/l  & [0.01, 0.08] & \cite{2000LKE}\\
    \hline
    $ K$ & Producer light-dependent carrying capacity \raggedright & varies mg C/l& [0.25, 2.0]& \cite{2000LKE}\\
    \hline 
    $ b$ & Producer max growth rate & 1.2/day& [0.75, 1.5]& \cite{2000LKE}\\
    \hline 
    $q$ & Producer min P:C  & 0.0038 mg P/mg C & [0.001, 0.005]& \cite{2000LKE}\\
    \hline 
    $e_j$ & Juvenile max production efficiency \raggedright & 0.5 & [0.2, 0.98] & assumed \\
    \hline 
    $e_a$ & Adult max production efficiency & 0.8 & [0.2, 0.98]& \cite{2000LKE}\\
    \hline 
    $\theta_j$ & Juvenile constant P:C ratio  & 0.025mg P/mg C & [0.02, 0.08] &assumed\\
    \hline 
    $\theta_a $ & Adult constant P:C ratio  & 0.03mg P/mg C & [0.02, 0.08]& \cite{2000LKE}\\
    \hline 
    $\delta_j$ &  Juvenile loss rate  & 0.06/day & [0.01, 0.3] &assumed\\
      \hline 
    $\delta_a$ &  Adult loss rate  & 0.08/day & [0.01, 0.3] &assumed \\
      \hline 
    $a_j$ &  Juvenile ingestion Half-saturation constant \raggedright& 0.25 mg C/l & [0.1, 0.5] &assumed\\
      \hline 
    $a_a$ & Adult ingestion Half-saturation constant \raggedright& 0.25 mg C/l & [0.1, 0.5]& \cite{2000LKE}\\
      \hline 
    $c_j$ & Juvenile max ingestion rate & 0.5/day & [0.1, 0.8] &assumed\\
      \hline 
    $c_a$ & Adult max ingestion rate & 0.81/day & [0.4, 1.2]& \cite{2000LKE}\\
   \hline
\end{tabular} 
\vspace{0.25cm}
   \caption{Stoichiometric co-limited stage-structure model \ref{eq:Model} parameter table.}
    \label{tab:params} 
\end{table}

\begin{figure}[H]
    \subfloat[$K=0.5$ mg C/l]{\includegraphics[width=.33\textwidth]{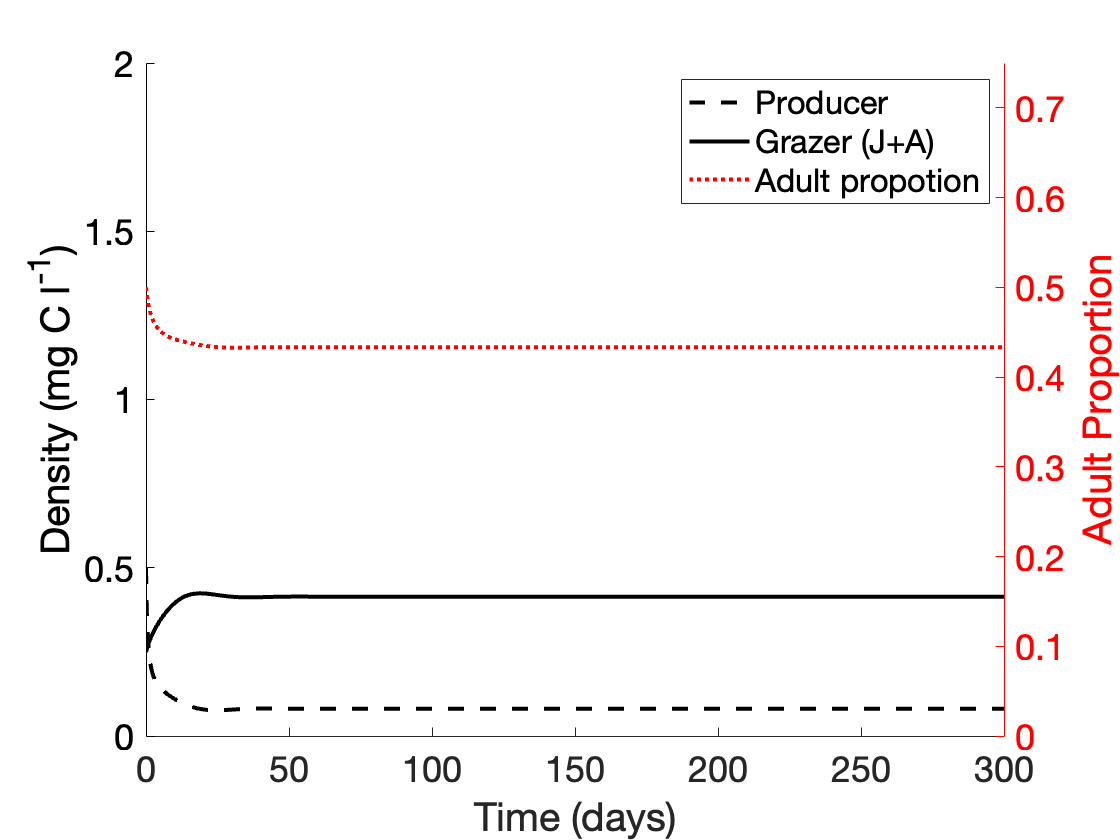}\label{fig:simsa}}     
     \subfloat[$K=1$ mg C/l]{\includegraphics[width=.33\textwidth]{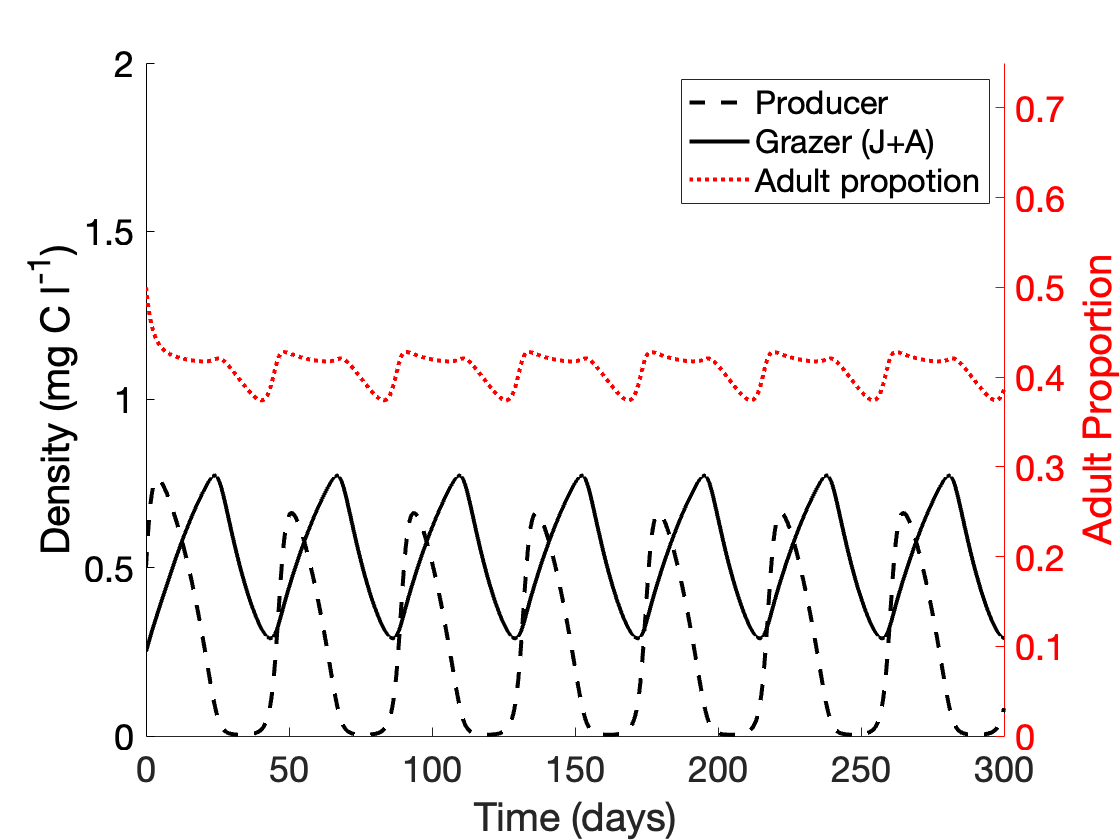}  \label{fig:simsb}}
    \subfloat[$K=2$ mg C/l]{\includegraphics[width=.33\textwidth]{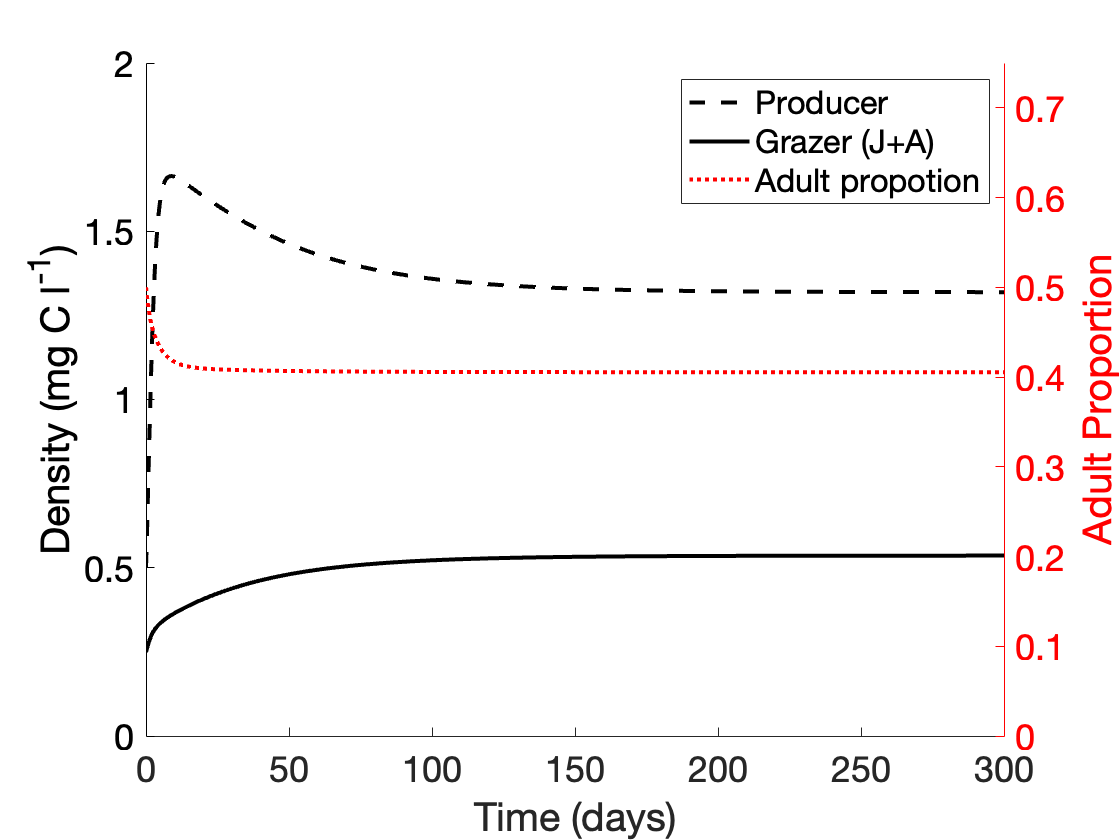}  \label{fig:simsc}}
    \caption{Numerical simulation of the model \eqref{eq:Model} showing population densities (black) and the proportion of C in the grazer population composed of adults (red), for parameter values listed Table~\ref{tab:params} and varying values for $K$ representing \textbf{(a)} low light levels $K=0.5$ mg C/l, \textbf{(b)} medium light levels $K=1$ mg C/l, and \textbf{(c)} high light levels $K=2$ mg C/l. Initial conditions are $x(0)=0.5$ mg C/l, $J(0)= 0.125$ mg C/l, and $A(0)=0.125$ mg C/l.}
    \label{fig:sims}
\end{figure}

\begin{figure}[H]
\centering
    \includegraphics[width=.45\textwidth]{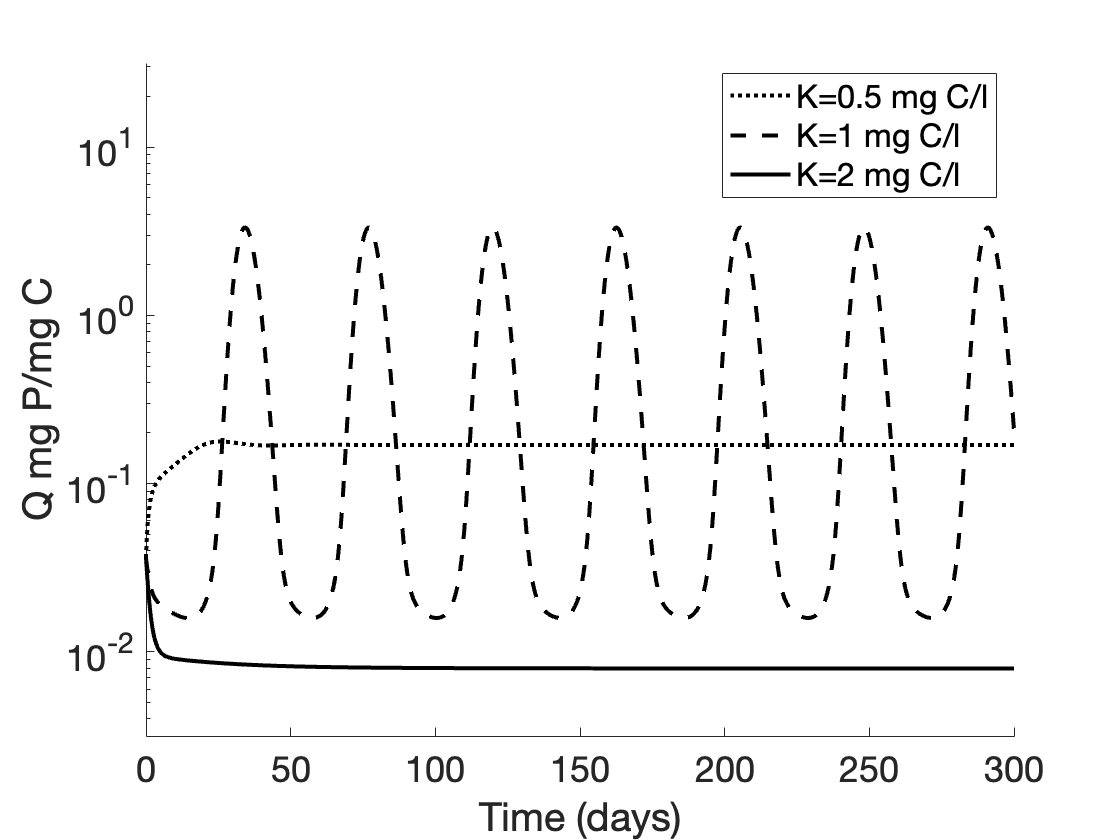}
    \caption{Numerical simulation of the model \eqref{eq:Model} showing $Q$, the variable P:C ratio of the producer, for parameter values listed Table~\ref{tab:params} and varying values for $K$.  Initial conditions are $x(0)=0.5$ mg C/l, $J(0)= 0.125$ mg C/l, and $A(0)=0.125$ mg C/l. Simulations correspond with those in Fig. \ref{fig:sims}.}
    \label{fig:simsQ}
\end{figure}

The light-dependent carrying capacity $K$ has been shown to be an important bifurcation parameter in several stoichiometric producer-grazer models \cite{2000LKE, 2015peace}. In Figure \ref{fig:sims}, we explore the influence $K$ has on the population dynamics, as well as the age distribution of juveniles and adult grazers with numerical simulations.
Under low light levels, the populations approach an interior equilibrium where both producer and grazer coexist (Fig. \ref{fig:simsa}). The low light conditions result in a relatively low producer density. Higher values of light yield oscillatory dynamics, where the populations coexist in stable limit cycles (Fig. \ref{fig:simsb}. Here, the proportion of carbon in the grazer population that composes adult biomass also oscillates. Interestingly, for even higher values of light the cycles collapse and solutions approach a new interior equilibrium (Fig. \ref{fig:simsc}. Here, the high light conditions result in a high producer density. Despite this high food density, the grazer equilibrium values are at similar values to the low light case observed in Fig. \ref{fig:simsa}. This is due to the stoichiometric constraints, as $Q$, the variable P:C ratio of the producer depends on a light levels, Fig. \ref{fig:simsQ}. While low light levels yield low producer density, the producers are relatively rich in P resulting in high food quality for grazers. On the other hand, high light levels yield high producer densities that have relatively low P content resulting in low quality food for grazers. Medium levels of light, $K=1$ mg C/l, yield oscillatory dynamics where food quality can vary across wide ranges.  

\subsection{Bifurcation Diagrams}
\begin{figure}[H]
\centering
\setlength{\tempwidth}{.3\linewidth}
\settoheight{\tempheight}{\includegraphics[width=\tempwidth]{Fig2a.png}}%
\columnname{Juvenile Density}\hfil
\columnname{Adult Density}\hfil
\columnname{Total Grazer Density}\\[1ex] 

\rowname{varying $\theta_j$}
\subfloat{\includegraphics[width=\tempwidth]{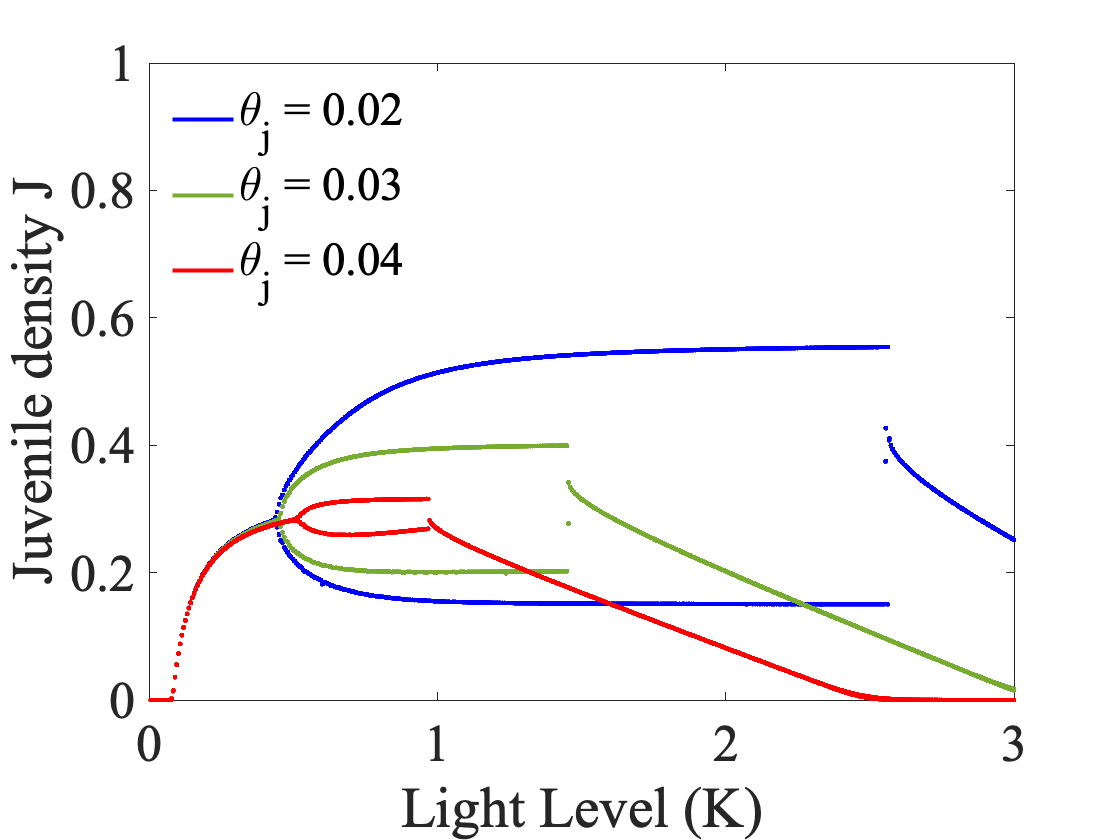}}\hfil
\subfloat{\includegraphics[width=\tempwidth]{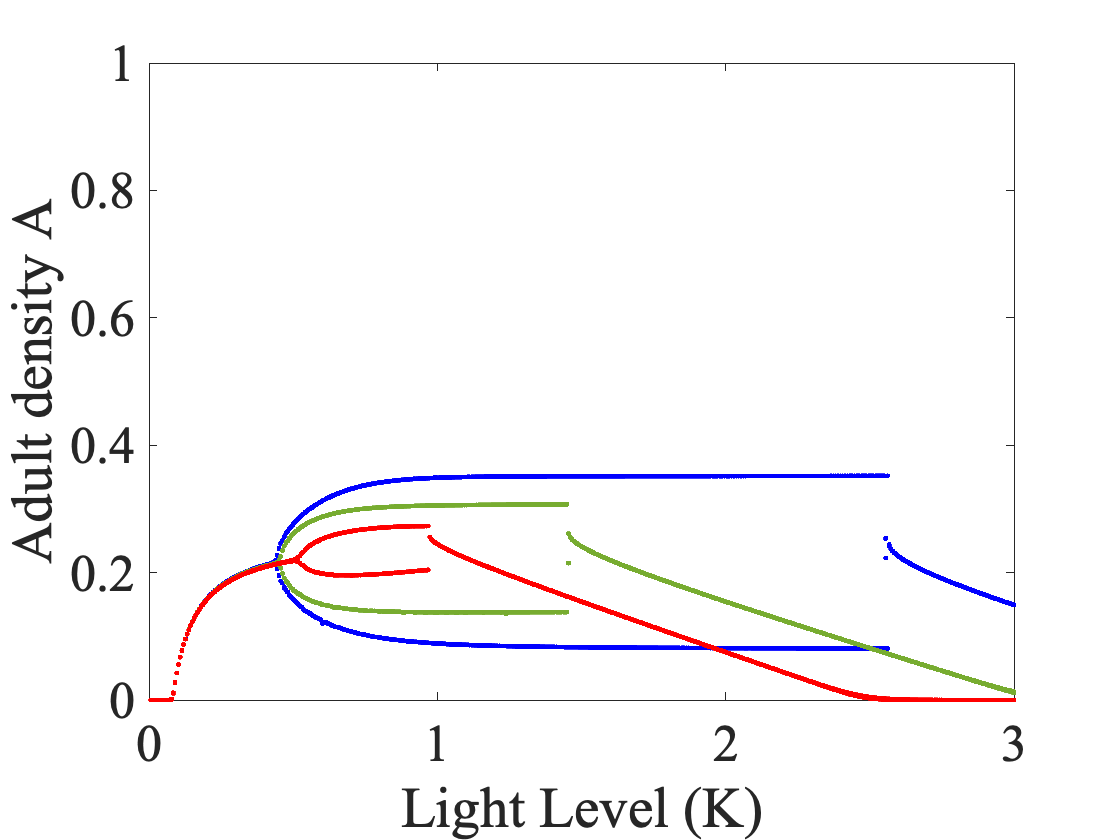}}\hfil
\subfloat{\includegraphics[width=\tempwidth]{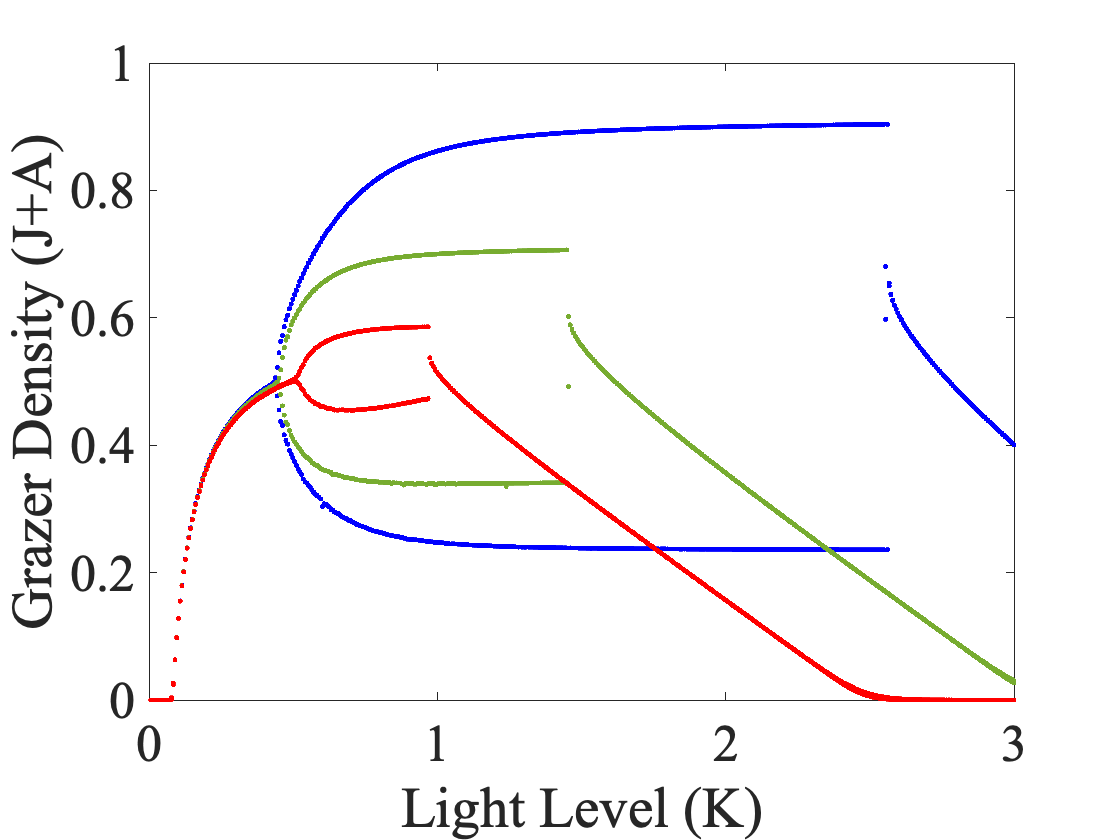}}\\[1ex] 

\rowname{varying $e_j$}
\subfloat{\includegraphics[width=\tempwidth]{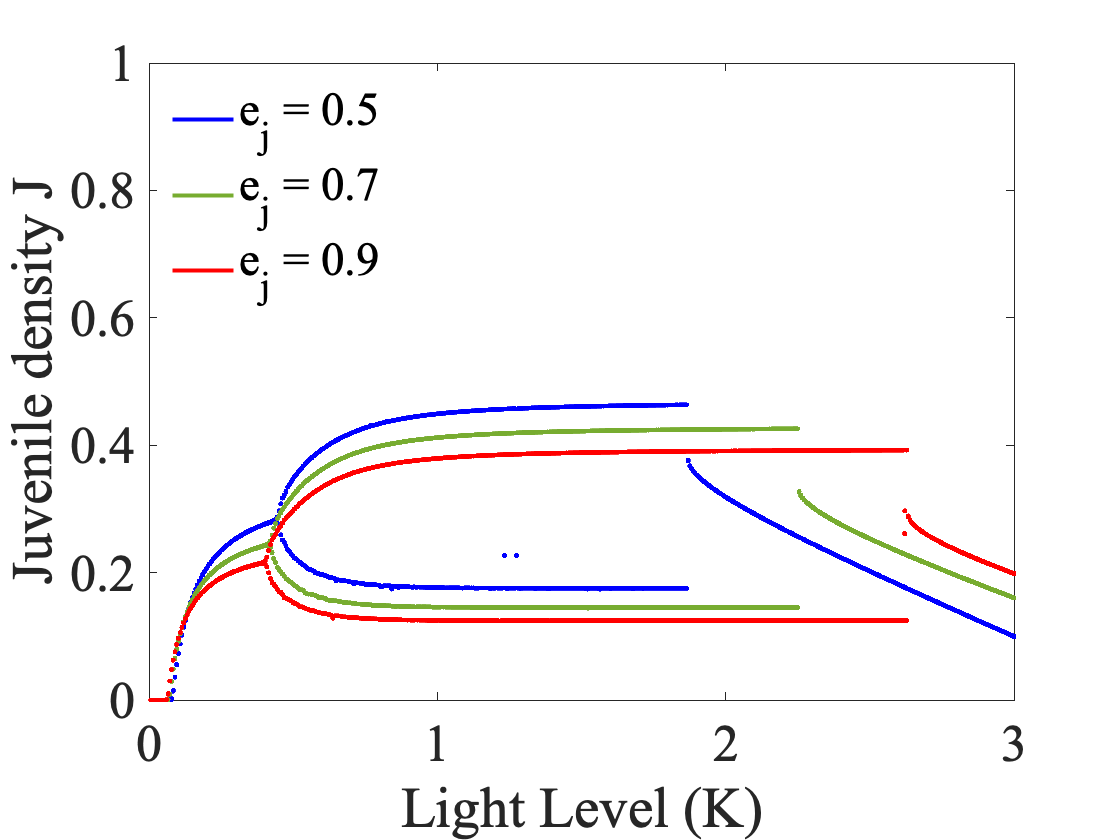}}\hfil
\subfloat{\includegraphics[width=\tempwidth]{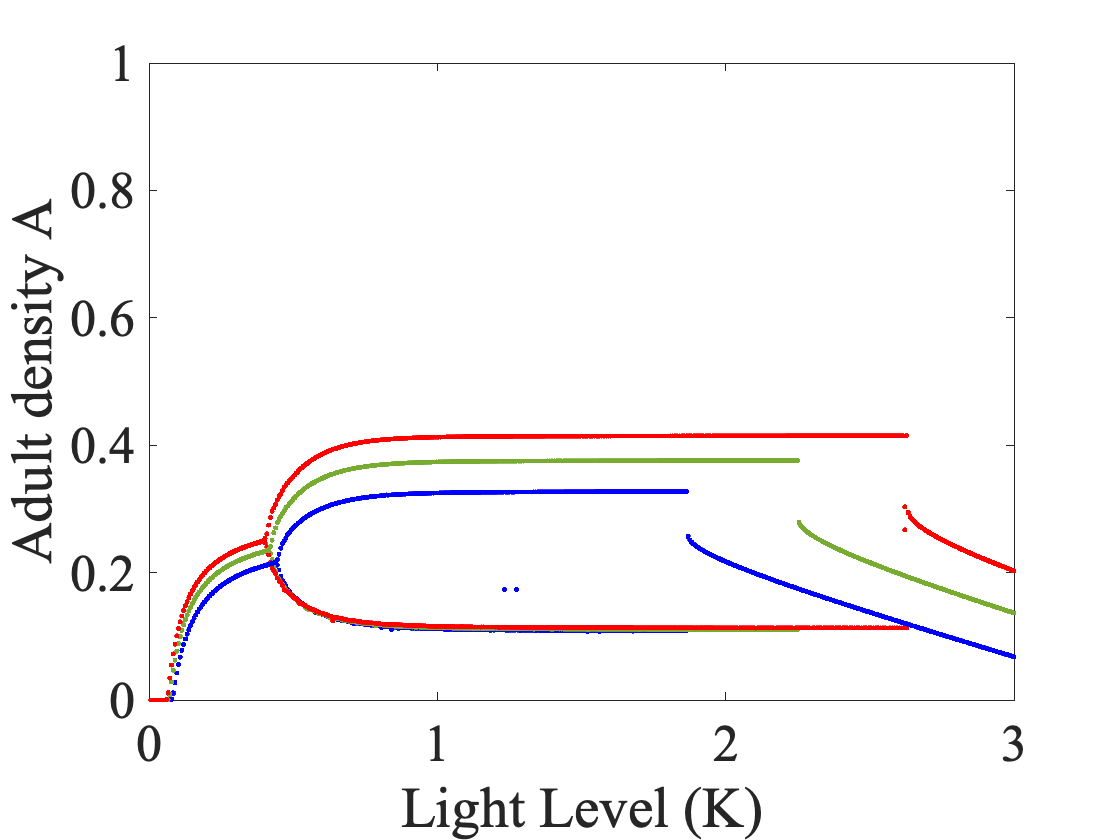}}\hfil
\subfloat{\includegraphics[width=\tempwidth]{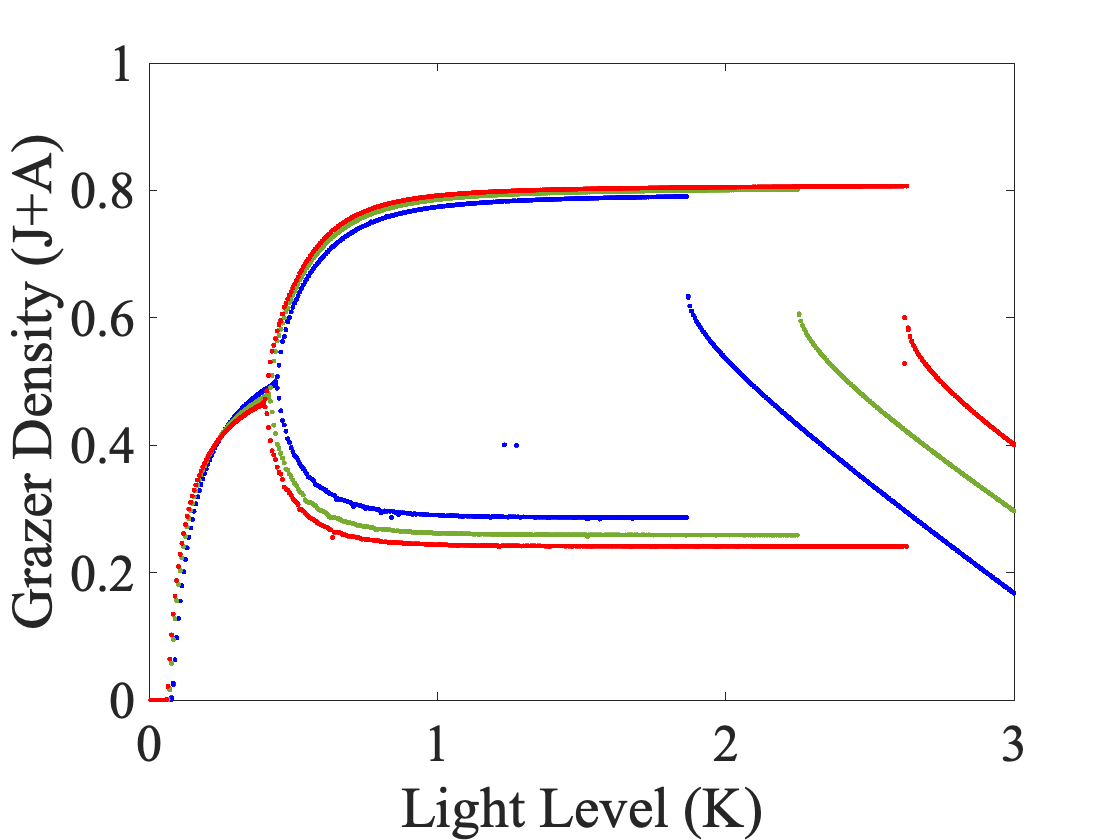}}\\[1ex]

\rowname{varying $c_j$}
\subfloat{\includegraphics[width=\tempwidth]{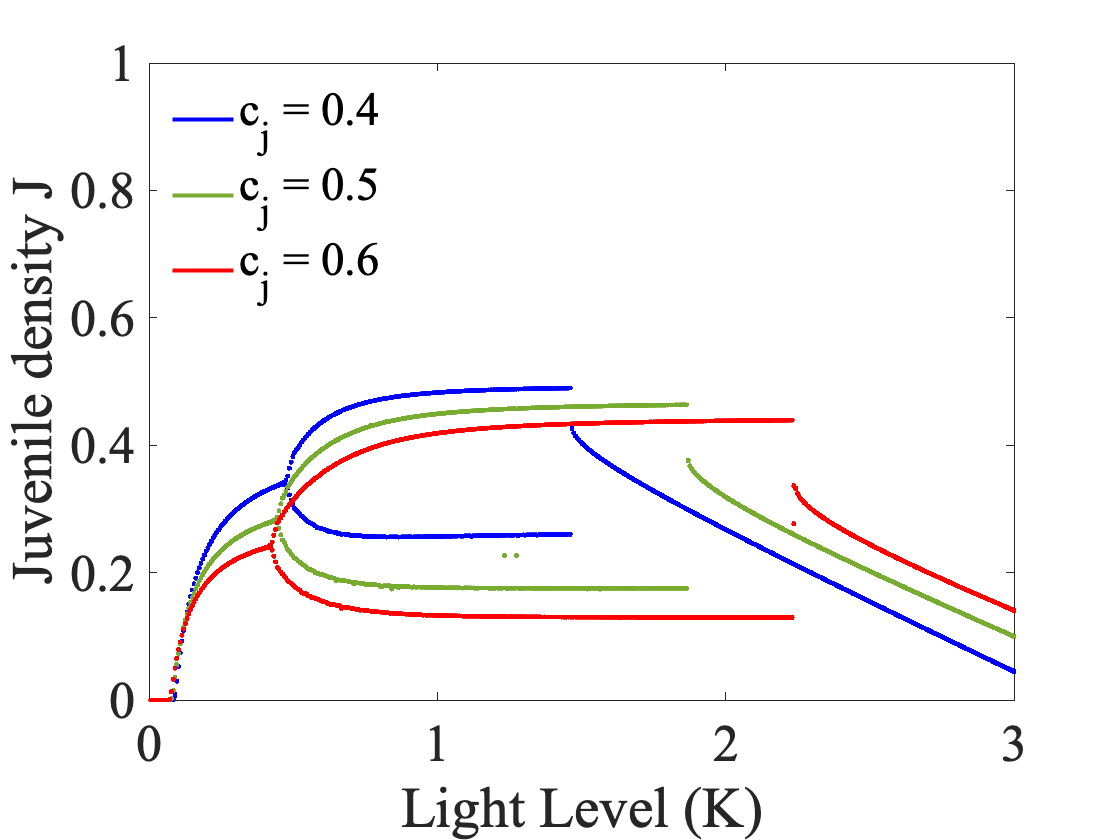}}\hfil
\subfloat{\includegraphics[width=\tempwidth]{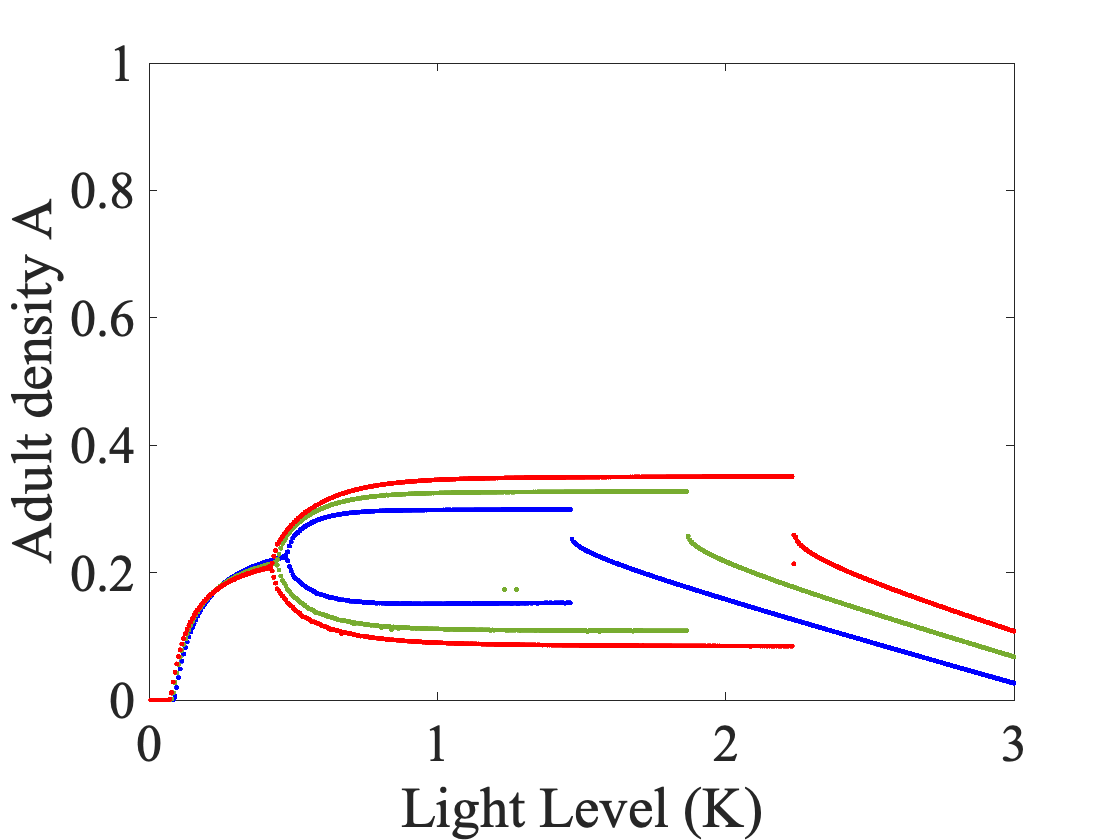}}\hfil
\subfloat{\includegraphics[width=\tempwidth]{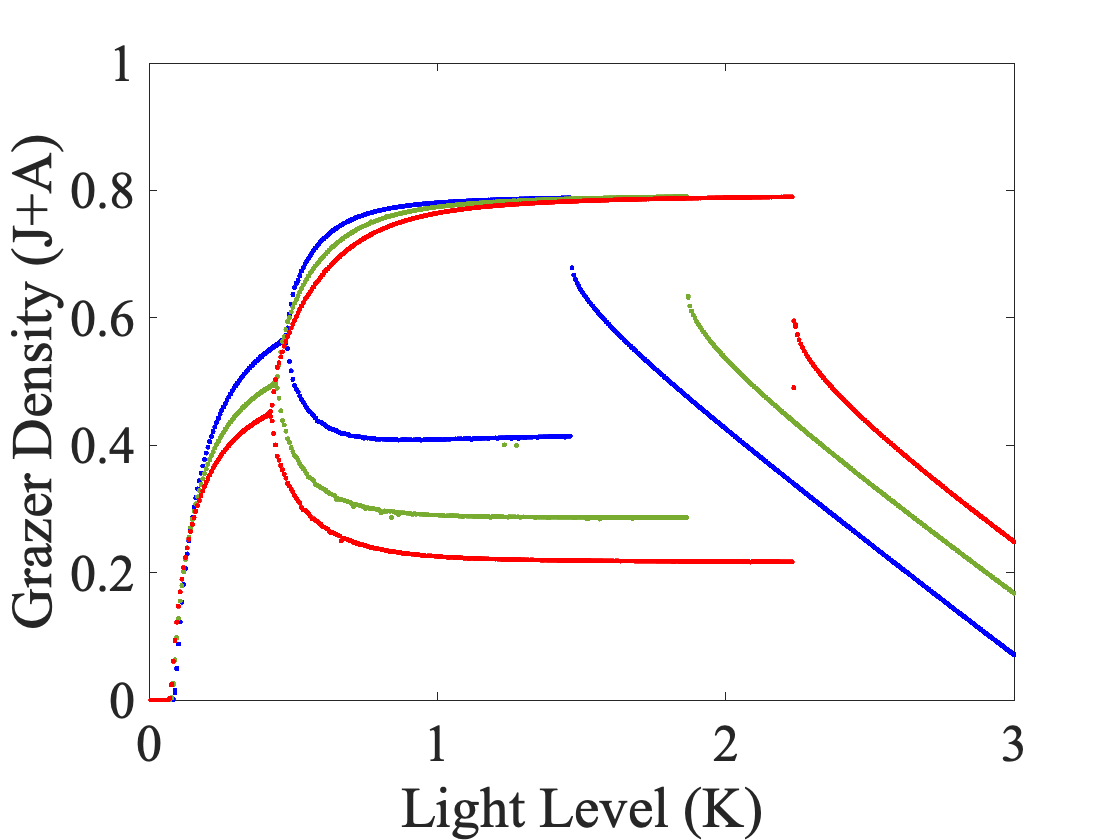}}\\[1ex]
 \caption{Bifurcation diagram of the model \eqref{eq:Model} using parameters in Table~\ref{tab:params} and bifurcation parameter $K$. These bifurcation diagrams show the long-term behavior ($t>1500$ days) of the juvenile density (left column), the adult density (middle column), and total grazer density (right column) for varying values of the juvenile P:C ratio $\theta_j$ (top row), juvenile maximum conversion efficiency $e_j$ (middle row), and juvenile maximum ingestion rate $c_j$ (bottom row).}
 \label{fig:Bifur}
 \end{figure}

We conducted numerical bifurcation analyses further exploring the dynamics for different light levels by varying the light-dependent producer carrying capacity $K$ (Fig. \ref{fig:Bifur}). In all cases the bifurcation structure is similar to that of previous stoichiometic producer-grazer models \cite{2000LKE, 2015peace}.  Grazers go extinct for very low values of $K$ where there is not enough producer to sustain their population, and  boundary equilibria $E_1$ or $E_2$ is stable. As $K$ increases an interior equilibrium gains stability and both species persist. The equilibrium grazer population density increases as $K$ increases until the systems reaches a Hopf bifurcation and limit cycles emerge. As $K$ continues to increase the limit cycles grow in amplitude then suddenly collapse. Here large amplitude limit cycles collapse at a saddle-node bifurcation and a new interior equilibrium gains stability. The grazer population then starts to decrease as $K$ increases and eventually dies out for very high values of $K$. These types of bifurcation dynamics observed in stoichiometric models have been robustly analyzed by Xie et al. 2018 \cite{xie2018complete}. 

Model \eqref{eq:Model} extends traditional stoichiometric models by incorporating stage structure. Here, we explore the important roles that juvenile specific parameters can play on the bifurcation dynamics of the system. Variations of juvenile P:C ratios, $\theta_j$, have profound effects on the limit cycles (Fig. \ref{fig:Bifur} top row), where higher $\theta_j$ values shorten the light ranges where limit cycles occur and decreases their amplitudes. Juveniles with higher $\theta_j$ have a higher P demand and are more sensitive to nutrient limitation. Under high light scenarios where nutrient limited growth dynamics are observed, grazer populations with higher $\theta_j$ values cause the grazer to die out at lower light levels. The ranges of $K$ exhibiting limit cycles are also influenced by juvenile maximum conversion efficiency $e_j$ (Fig. \ref{fig:Bifur} middle row) and their maximum ingestion rates $c_j$ (Fig. \ref{fig:Bifur} bottom row).  Both $e_j$ and $c_j$ appear in the juvenile maturation rate, and changes in their values modify the population density of juveniles and of adults.  Variations in $e_j$ do not have large affects on the amplitudes of the cycles, but do produce a shift in the location of the saddle-node bifurcation and the collapse of the cycles. Variations in $c_j$ affect the amplitudes of the cycles of the total grazer population densities (Fig. \ref{fig:Bifur} right column, bottom row) and bifurcation points of the populations.  For low light conditions, increases in $e_j$ and $c_j$ decreases juvenile densities and their proportion of the grazer populations. However, in high light conditions increases in $e_j$ and $c_j$ increase juvenile densities and their proportion of the grazer populations.  Biologically, increases to the maturation rate will shorten the juvenile stage, which can decrease juvenile population densities. On the other hand, increased maturation rates will increase adult populations leading to higher reproduction, which can increase juvenile population densities. 

\begin{figure}[H]
\centering
\subfloat[Juvenile Density, $K=0.5$ mg C/l]{\includegraphics[width=.45\textwidth]{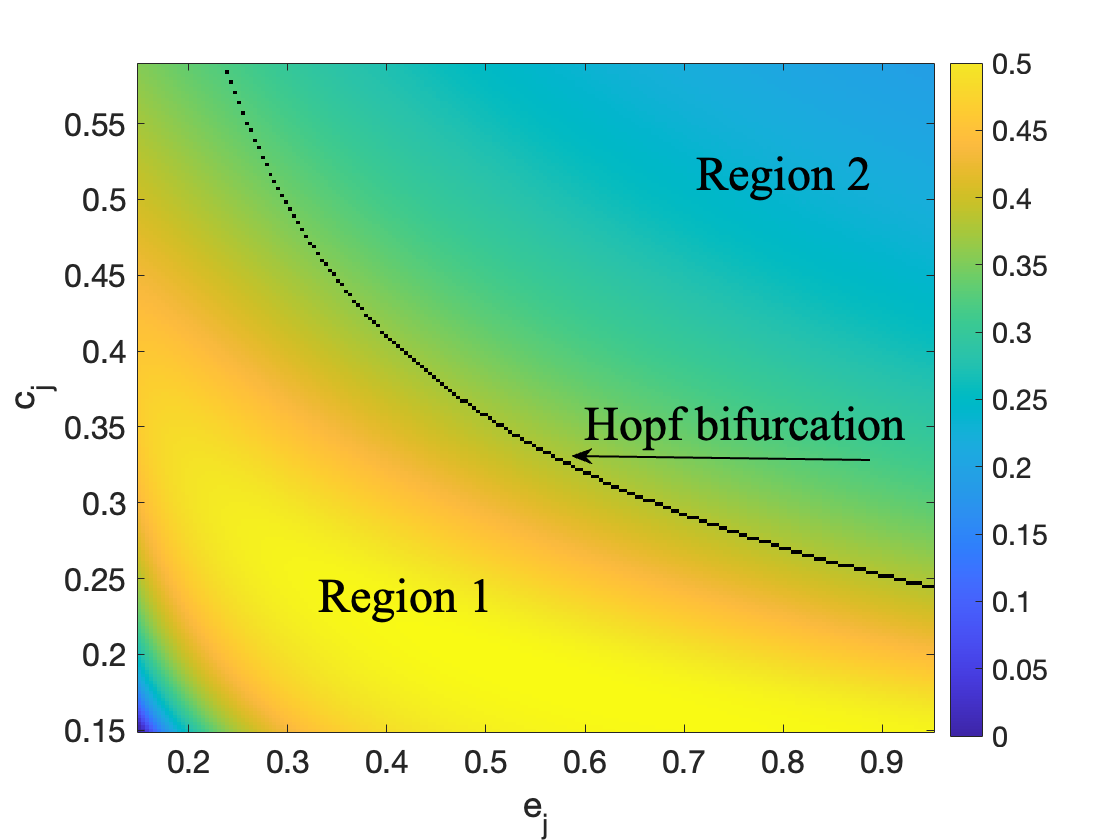}}
 \subfloat[Adult Density, $K=0.5$ mg C/l]{\includegraphics[width=.45\textwidth]{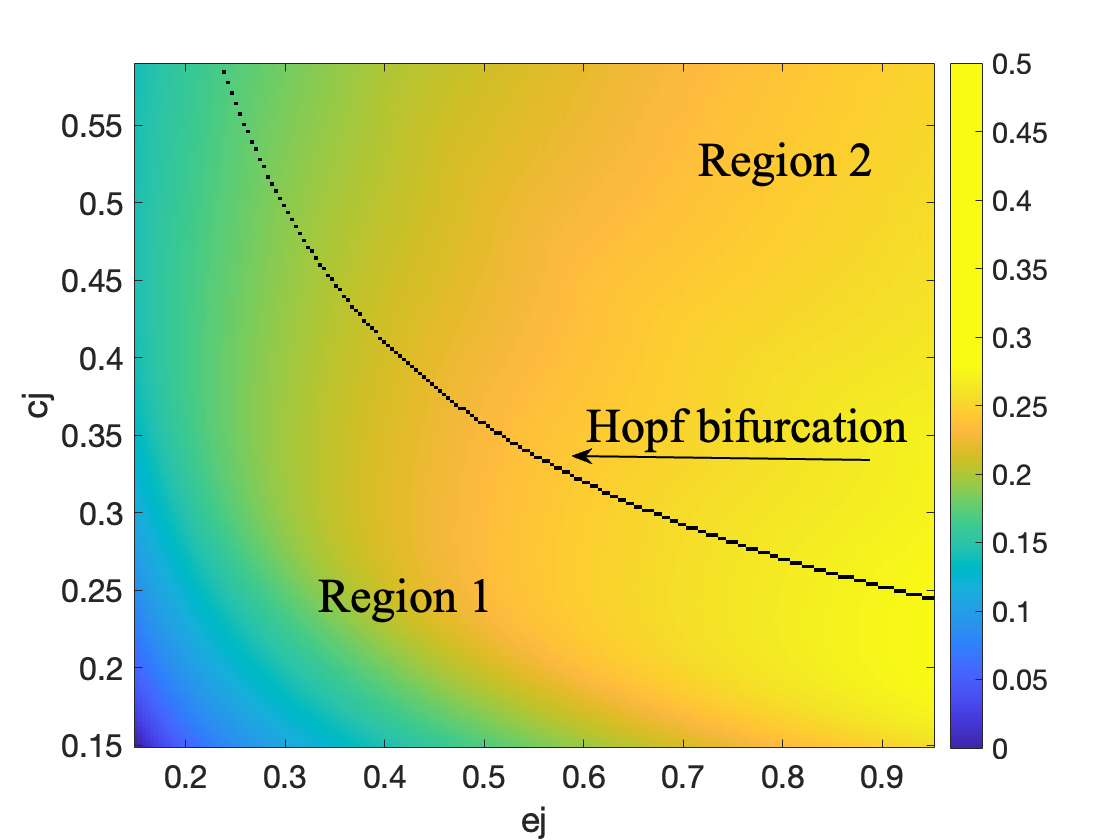}}\\
\subfloat[Juvenile Density, $K=1$ mg C/l]{\includegraphics[width=.45\textwidth]{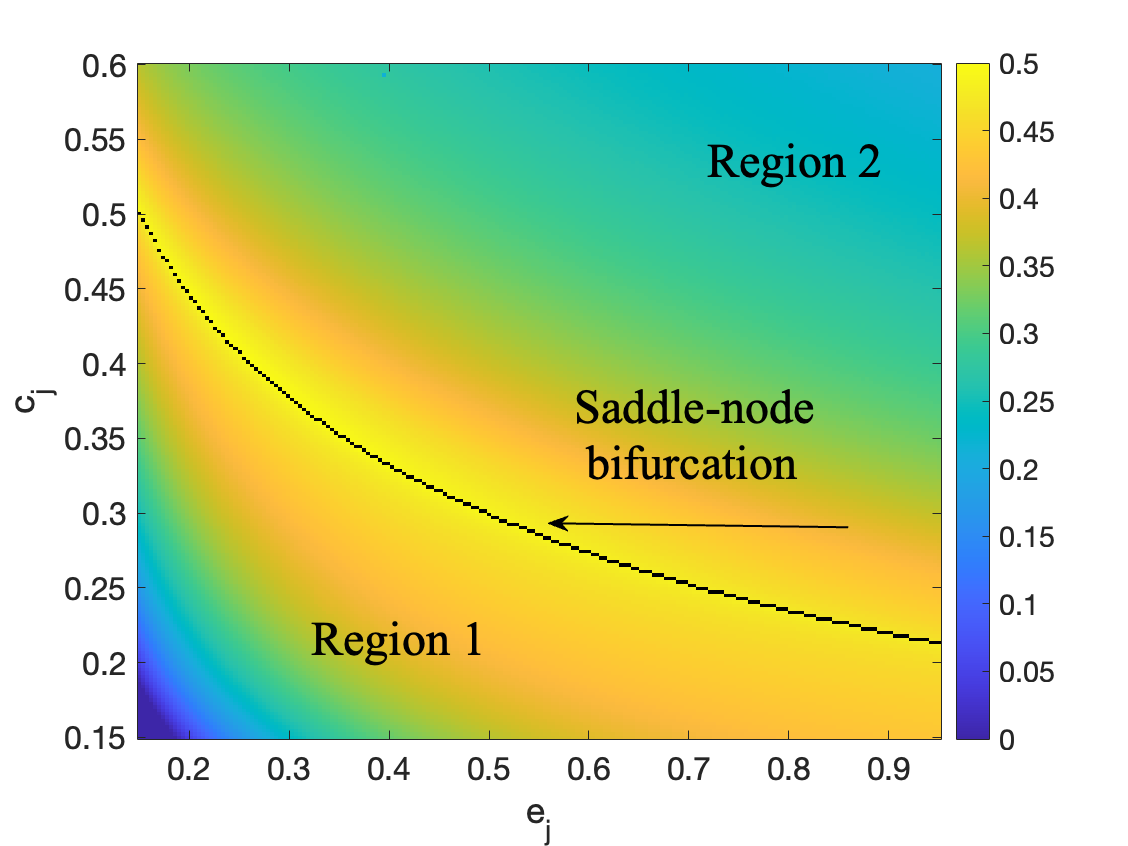}}
 \subfloat[Adult Density, $K=1$ mg C/l]{\includegraphics[width=.45\textwidth]{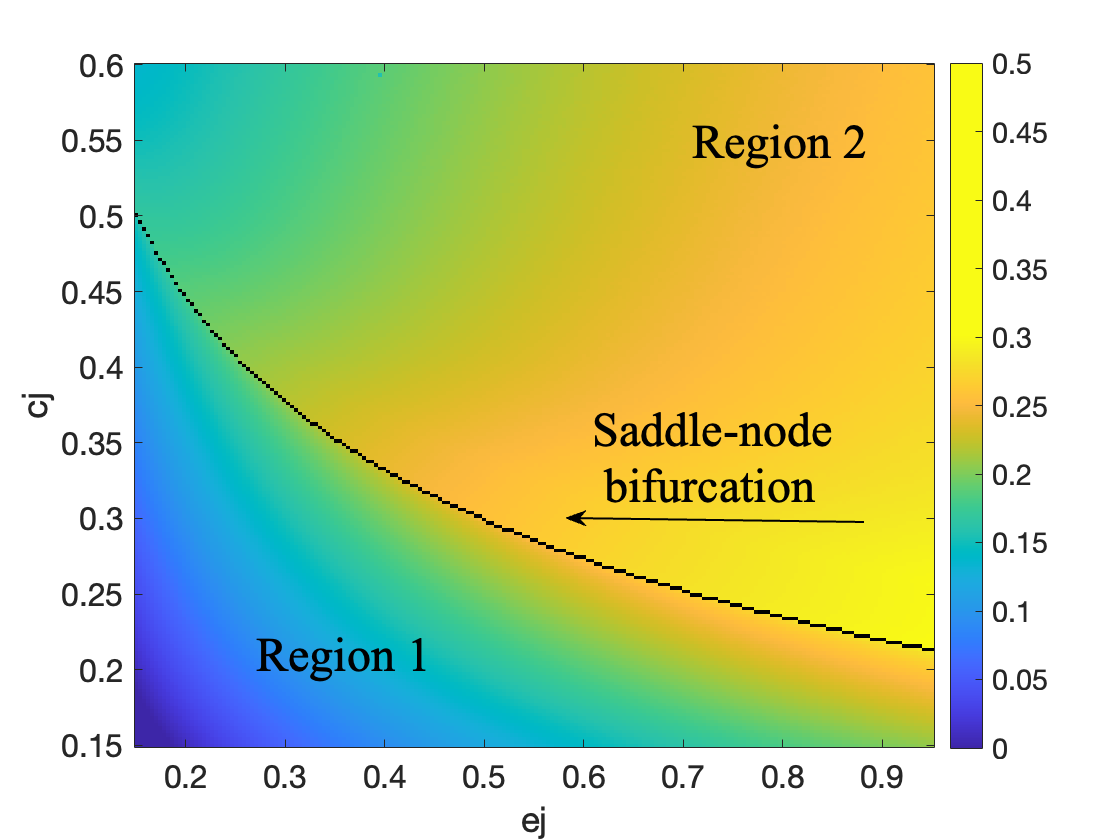}}
 \vspace{0.25cm}
 \caption{Two parameter bifurcation showcasing long term dynamics ( $>5000$ days) of (a) Juvenile and (b) Adult population densities when light-dependent producer carrying capacity $K=0.5$ mg C/l; and (c) Juvenile and (d) Adult population densities when light-dependent producer carrying capacity $K=1$ mg C/l for varying values of juvenile maximum ingestion rate $c_j$ and juvenile maximum conversion efficiency $e_j$. Black curves separates regions where long term dynamics approach equilibria and limit cycles. The black curve in (a) and (b) corresponds with the Hopf bifurcation. The black curve in (c) and (d) corresponds with the collapse of high amplitude cycles at the saddle-node bifurcation. Region 1 exhibits equilibria dynamics and Region 2 exhibits limit cycles, where the average density of the limit cycles is shown in the heatmap. }
 \label{fig:sims_heat}
 \end{figure}
We further explore the roles of $e_j$ and $c_j$ with two-parameter bifurcation analyses (Fig. \ref{fig:sims_heat}). These figures show the average population densities exhibited in long-term dynamics as heatmaps across  $e_j$ and $c_j$ space, as well as bifurcation curves which divide the parameter space into regions with stable equilibria and stable limit cycles.  Shifts in $e_j$ and $c_j$ affects the Hopf bifurcation point at low light levels (Fig. \ref{fig:sims_heat} (a)(b)), as well as the saddle-node bifurcation point that aligns with the collapse of the cycles at higher light levels (Fig. \ref{fig:sims_heat} (c)(d)).

\subsection{Sensitivity analysis}

\begin{figure}[H]
\centering
\subfloat[]
{ \includegraphics[width=1\linewidth]{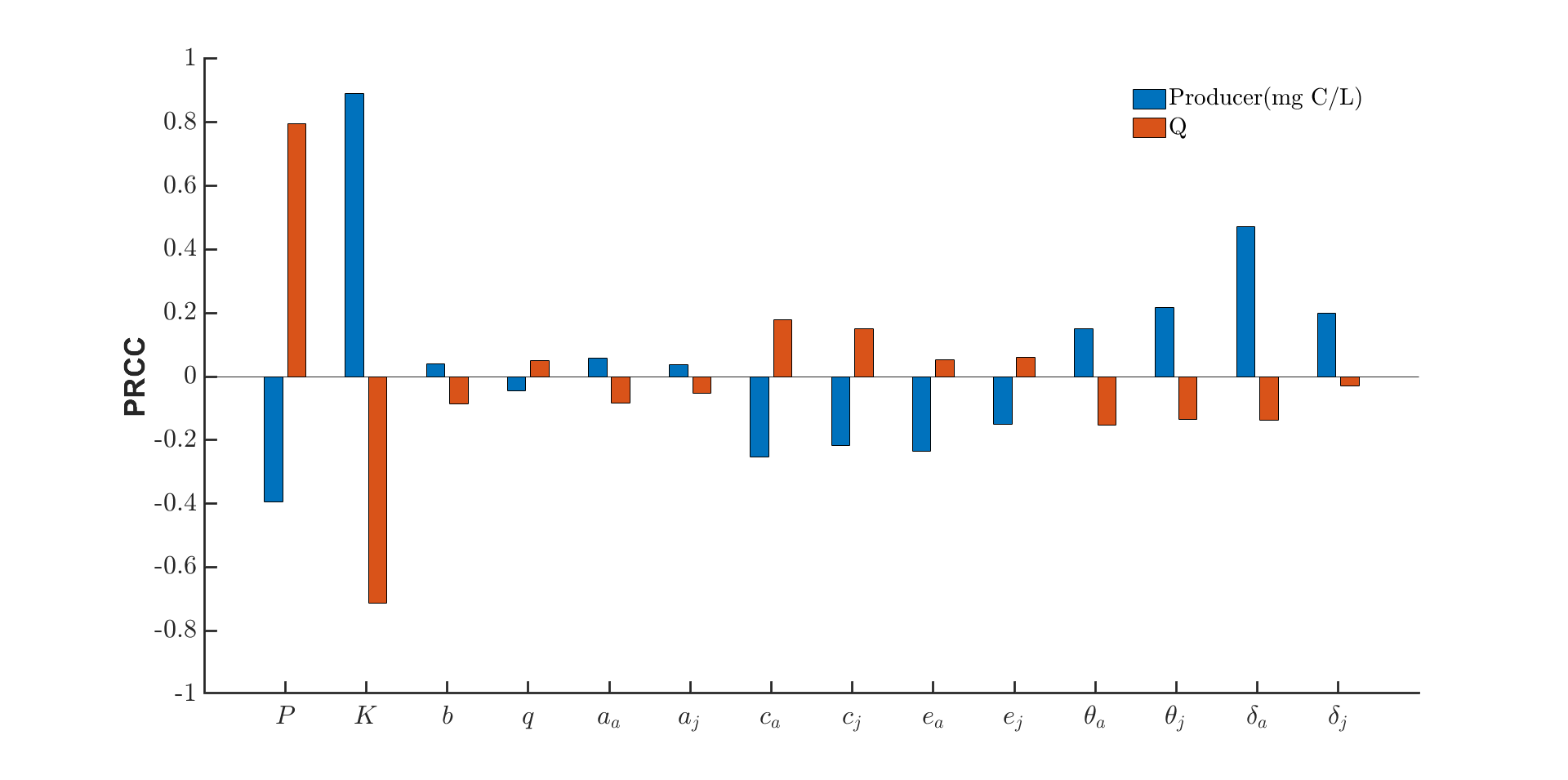}} \\
\subfloat[]
{ \includegraphics[width=1\linewidth]{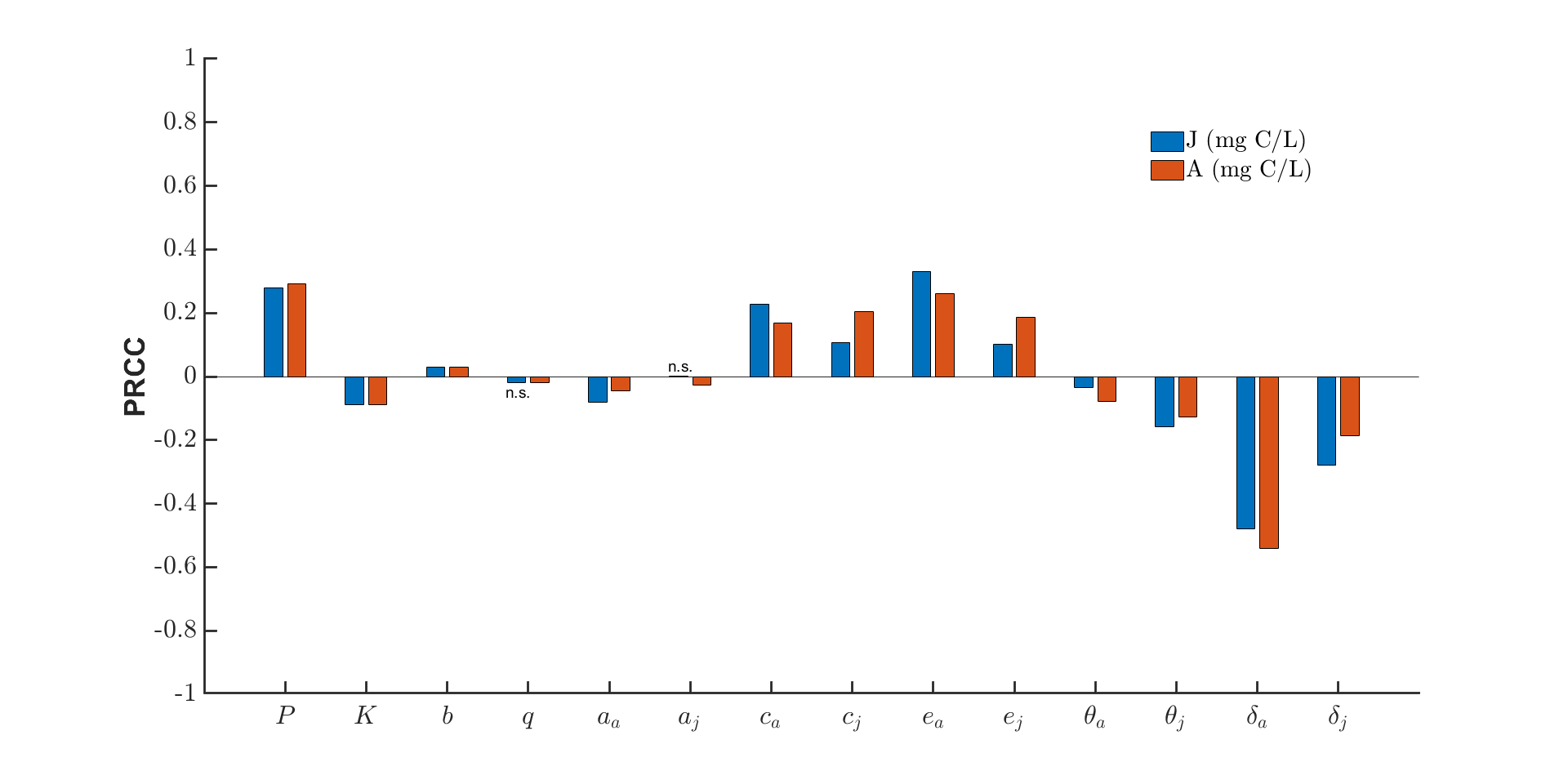}}
 \hspace{0.1cm}
  \caption{PRCC values for output measures: (a) average producer density (mg C/L) and P:C ratio of the producer and (b) average juvenile and adult densities (mg C/L), resulting from a sensitivity analysis of parameter ranges listed in Table~\ref{tab:params}. PRCC values range from -1 to 1, where negative values indicate an inverse relationship between the parameter and the output measure, and positive values suggest a direct positive impact on the output. Values highest in magnitude are the most influential.  Values marked n.s. are non-significant (statistical significance  $p > 0.05$). }
\label{fig:PRCC}
\end{figure}

We investigate the sensitivity of the stoichiometric co-limited stage-structure model \eqref{eq:Model} parameters using Latin Hypercube Sampling (LHS) in conjunction with the Partial Rank Correlation Coefficient (PRCC) technique to conduct a global sensitivity analysis and identify key parameters strongly correlated with the model's output. In this model, all 14 parameters were treated as uncertain, and we performed LHS with 10,000 simulations. Each uncertain parameter was assigned a uniform distribution with the ranges in Table \ref{tab:params} to determine the corresponding probability density function. We considered four outcome measures: 1.) average producer density (mg C/L), 2.) average juvenile density (mg C/L), 3.) average adult density (mg C/L), and 4.) P:C ratio of the producer. The PRCC is an appropriate statistical measure if parameters are monotonically related to output measures \cite{2008Marino}, and therefore we checked the relationship between each parameter individually and the output measures, (Figs. \ref{fig:Monotonicity_1} and \ref{fig:Monotonicity_2}). Many regions showed monotonic trends and we computed the PRCC as a sensitivity index (Fig \ref{fig:PRCC}). However regions of non-monotonicity are discussed in \ref{app:monotonic_split}.  Additionally, following Marino et al. \cite{2008Marino}, we performed a z-test on transformed PRCC values to identify significant parameters based on their relative sensitivity. The result of the z-test showed that, in general, parameters with PRCC values of greater magnitude exerted a strong influence on the output measures. 

According to the PRCC values, the environmental parameters of total phosphorus in the system $P$ and light-dependent carrying capacity $K$ significantly influence the model, as expected (Fig. \ref{fig:PRCC}).  Additionally, the stage-specific parameters $(c_j,c_a, e_j, e_a, \theta_j, \theta_a, \delta_j, \delta_a)$  strongly affect both average juvenile and adult densities (Fig. \ref{fig:PRCC}(b)). Here, increases in juvenile maximum ingestion rate $c_j$ and juvenile maximum production efficiency $e_j$ increase juvenile and adult densities.  Increases in juvenile P:C ratio $\theta_j$ and juvenile loss rate $\delta_j$ decrease densities. These results provide deeper insight into stage-specific processes within the model.

\section{Discussion}
\label{sec:discussion}

Ecological stoichiometry \cite{sterner2017ecological} considers the effects of both food quantity and food quality on population dynamics. While many mathematical models consider the theory of ecological stoichiometry to investigate the effects of nutrient changes in the environment \cite{2000LKE, 2015peace, 2008Wang}, few consider the life stages of the grazer populations and how nutrient limitation affects stage-specific processes. Biological studies show that juveniles may be  especially affected by nutrient limitation due to their higher growth rate and that nutrient poor food can have strong effects on reproduction output \cite{andersen2004stoichiometry}. We develop a stoichiometric co-limited stage-structure producer-grazer model \eqref{eq:Model} to investigate the effects of stoichiometry-dependent organismal stage structures under nutrient constraints. 

Stoichiometric constraints incorporated in the model showcase similar bifurcation structures to previously developed non-stage stoichiometric modes \cite{2000LKE}, where changes in light levels shift population dynamics from stable extinction equilibria to stable interior equilibria and stable limit cycles (Figures \ref{fig:sims}, \ref{fig:simsQ}, and \ref{fig:Bifur}). 
While light levels ($K$) and environmental phosphorus loads ($P$) have the most influence on model output, variations of stage-specific parameters impact populations dynamics and the variable P:C ratios of the producer (Fig. \ref{fig:PRCC}).  Indeed, model analyses showcase that variation in stage-specific parameters can lead to differences in the population dynamics. Specifically, juveniles with higher constant P:C ratios ($\theta_j$) have a shorter range of light levels that exhibit smaller-amplitude oscillations and these populations reach extinction at lower light levels compared to those with lower P:C ratios. Parameters affecting the maturation rate ($e_j$, $c_j$) also affect population dynamics; juveniles with higher maximum production efficiency and maximum ingestion rate have lower juvenile densities at low light levels, longer ranges of light levels that exhibit oscillatory population dynamics, and higher light levels are required for extinction conditions. 

In general, the influences of stage-specific parameters have higher effects on the population dynamics at higher light levels than at lower light levels, where we see the collapse of the limits cycles at the saddle-node bifurcation (Fig. \ref{fig:Bifur}). Higher light conditions correspond with a higher light-dependent carrying capacity for the algae population and lead to high quantity, but low quality food. The influence of stoichiometric constraints is more pronounced in these environments where algae exhibit lower P:C ratio \cite{sterner2017ecological}. 

While this model is an important first step to investigating the relationship between nutrient limitation and stage-specific processes, future work can consider a delay differential equation model to be more biologically realistic. Our stoichiometric co-limited stage-structure model \eqref{eq:Model} assumes an instantaneous maturation, whereas a delay differential equation model can incorporate a maturation rate. Realistically, this maturation rate can depend on the quantity as well as quality of the producer, which leads to a stage-dependent maturation delay. McCauley et al. \cite{2008McCauley} developed a stage-dependent maturation delay differential equations model allowing maturation rates to depend on food quantity and extending this framework stoichiometrically would be insightful. Future models can also consider the consequences of nutrient excess \cite{2014peace, elser2016life} and/or size-structure \cite{de1992studying} to allow for individual growth outside of maturation and reproduction. 

\section*{Acknowledgments}
RE and TA were supported by NSF DMS-2322103. DPD, AP, and GDM were supported by NSF DMS-2322102. JL was supported by NSF DMS-2322104. 


\appendix

\section{Latin Hypercube and Partial Rank Correlation
Coefficient}
\label{app:monotonic_split}
\begin{figure}[H]
\subfloat[Average producer density (mg C/L)]{\includegraphics[width=1\linewidth]{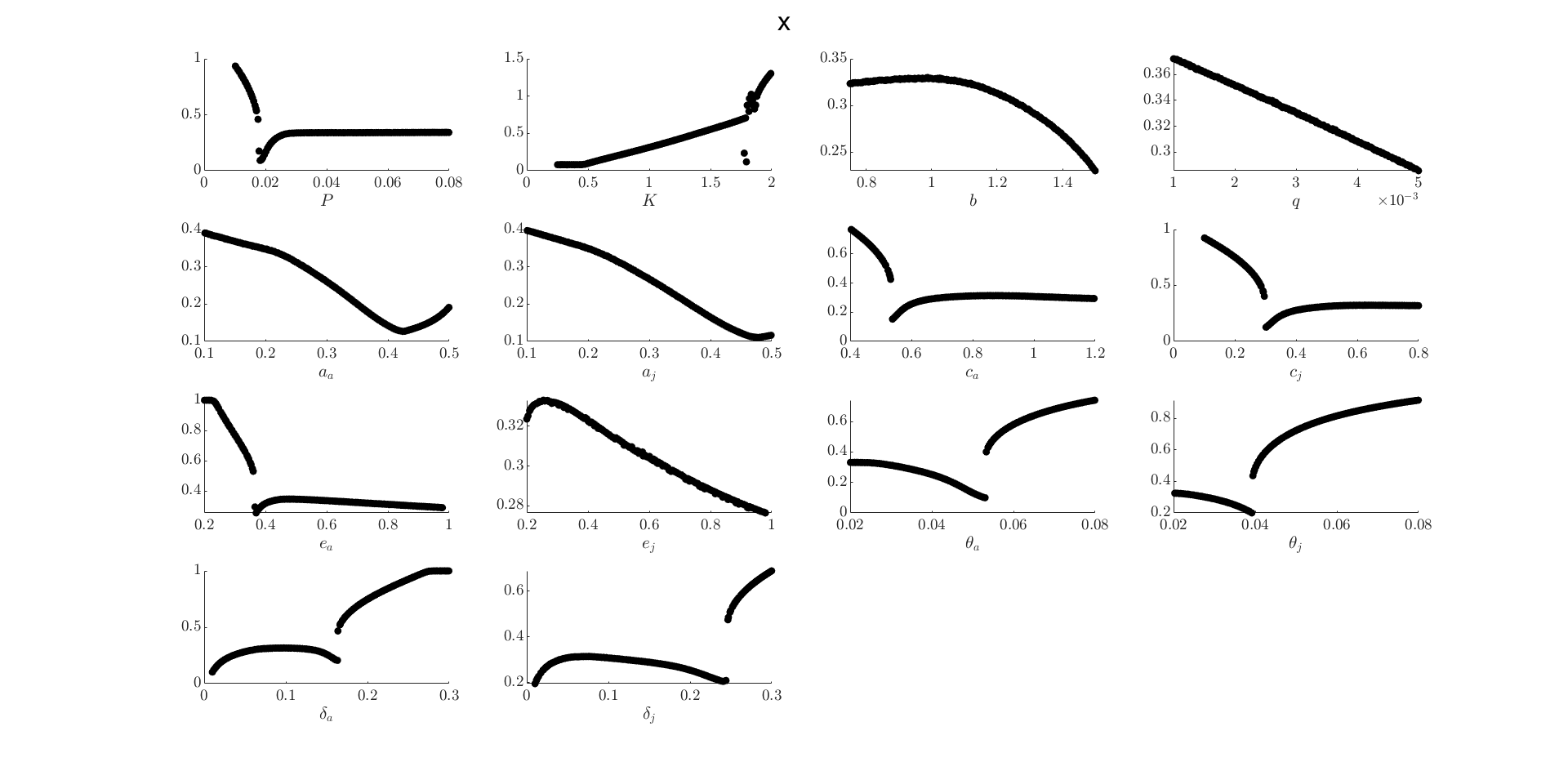}} \\
\subfloat[Average juvenile density (mg C/L)]{\includegraphics[width=1\linewidth]{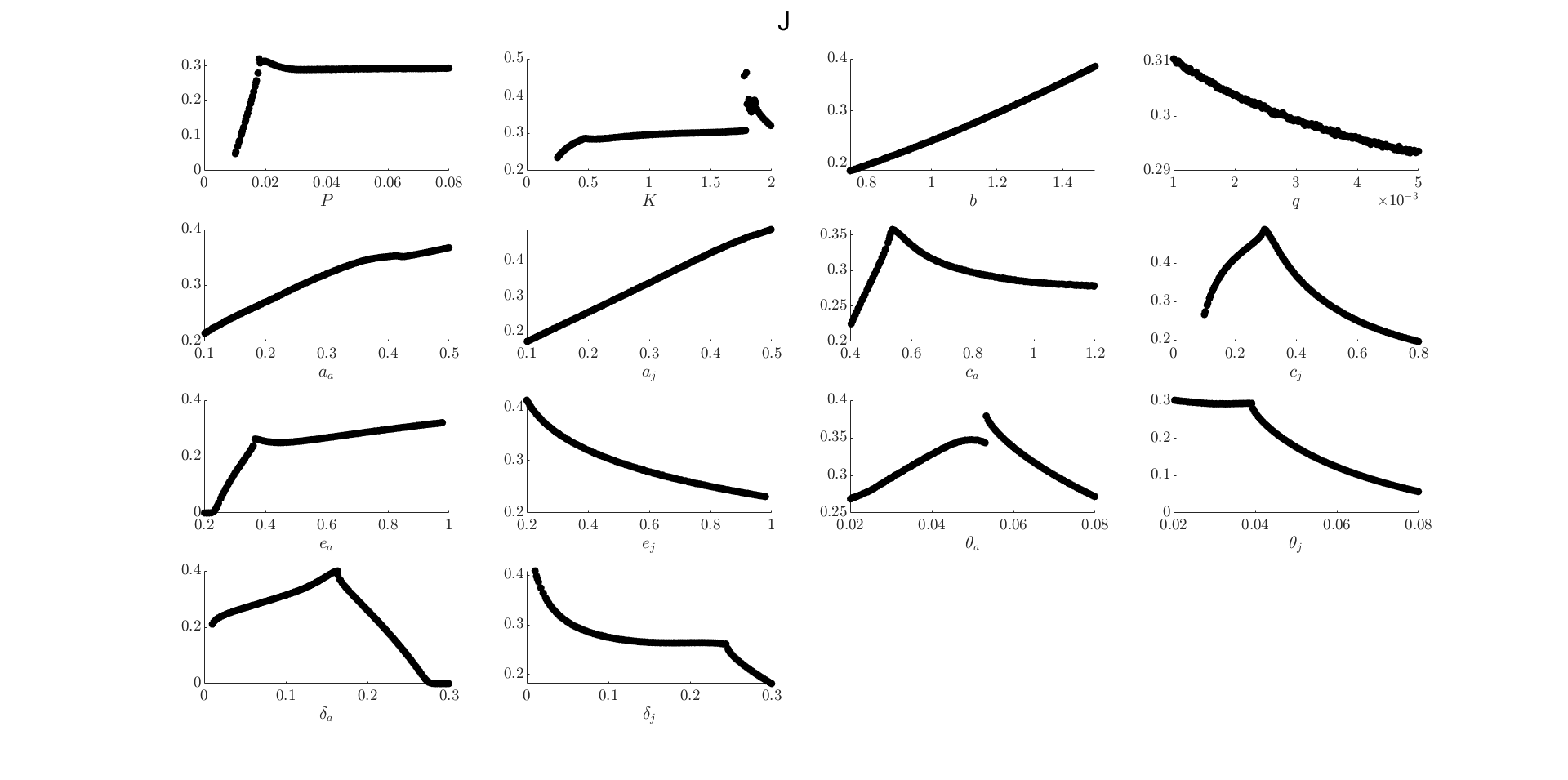}} 
  \caption{Monotonicity plots for (a) average producer density (mg C/L) and (b) average juvenile density (mg C/L) by using parameter values listed in Table~\ref{tab:params}.}
 \label{fig:Monotonicity_1}
 \end{figure}

\begin{figure}[H]
\subfloat[Average adult density (mg C/L)]{\includegraphics[width=1\linewidth]{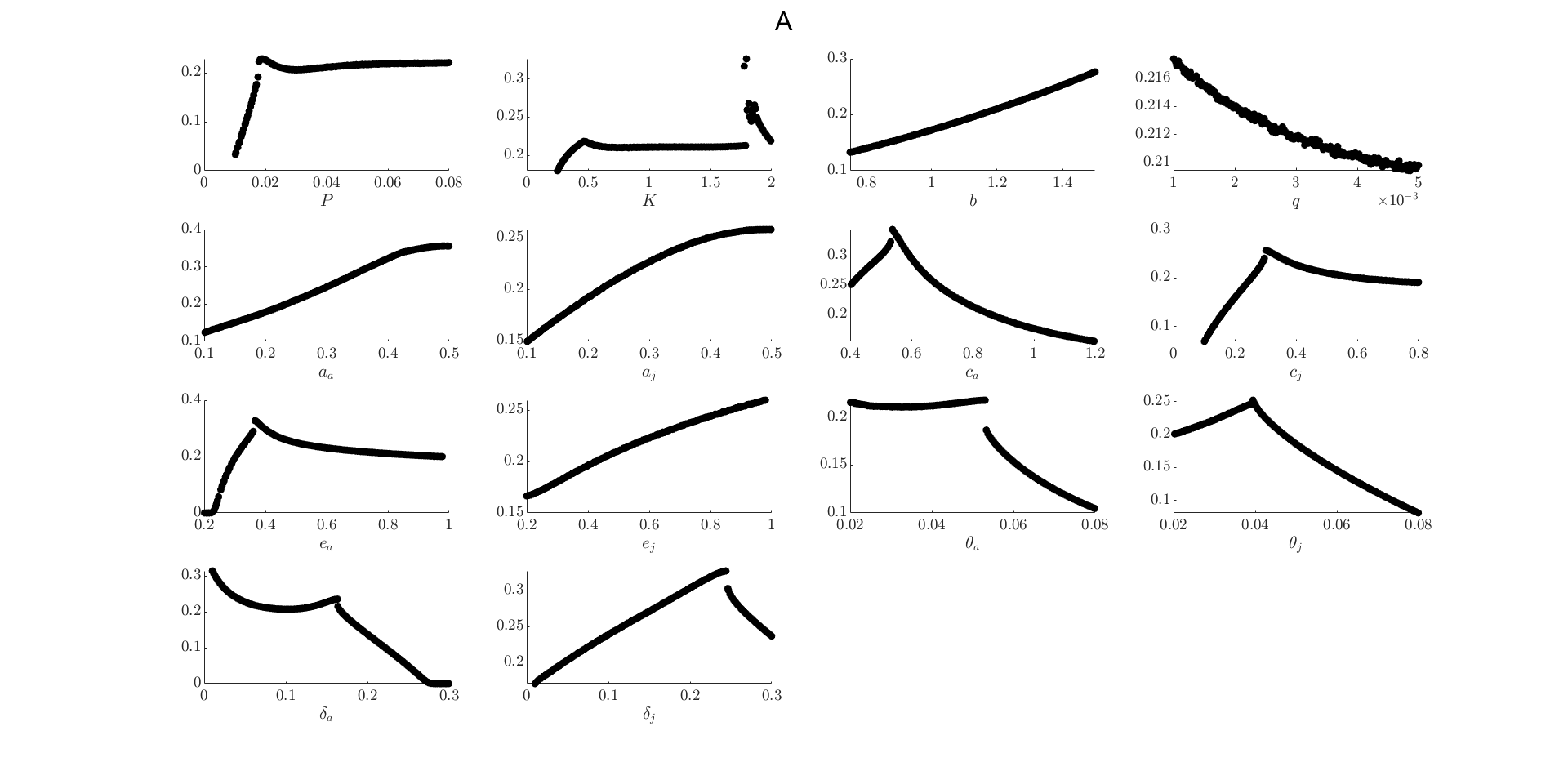}} \\
\subfloat[P:C ratio of the producer]{\includegraphics[width=1\linewidth]{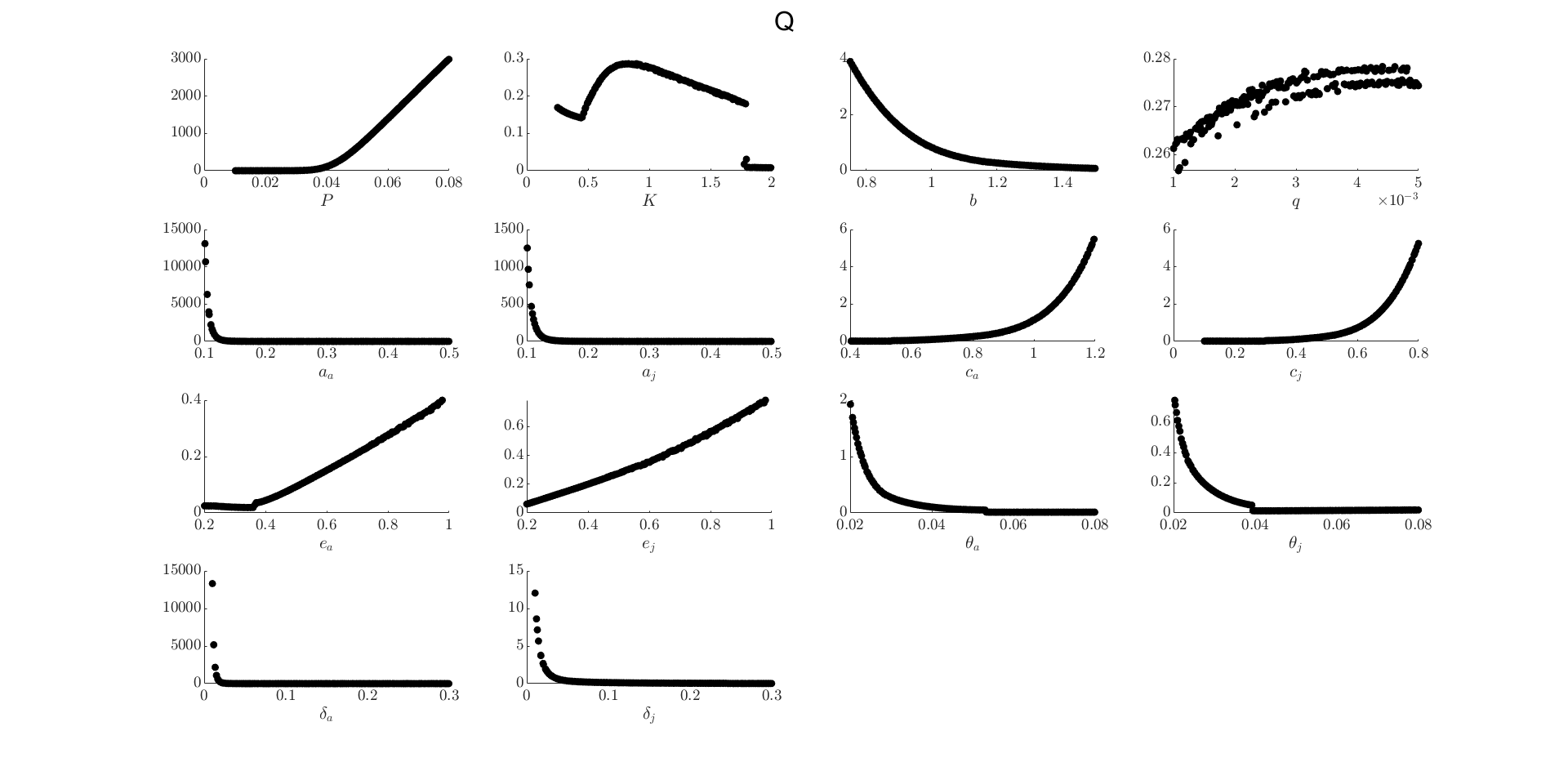}} 
  \caption{Monotonicity plots for (a) average adult density (mg C/L) and (b) P:C ratio of the producer by using parameter values listed in Table~\ref{tab:params}.}
 \label{fig:Monotonicity_2}
 \end{figure}

\begin{figure}[H]
\centering

\begin{minipage}{0.49\textwidth}
    \centering
    \subfloat[$c_j$ range 0.1-0.3 (effect on x and Q)]{\includegraphics[width=\textwidth]{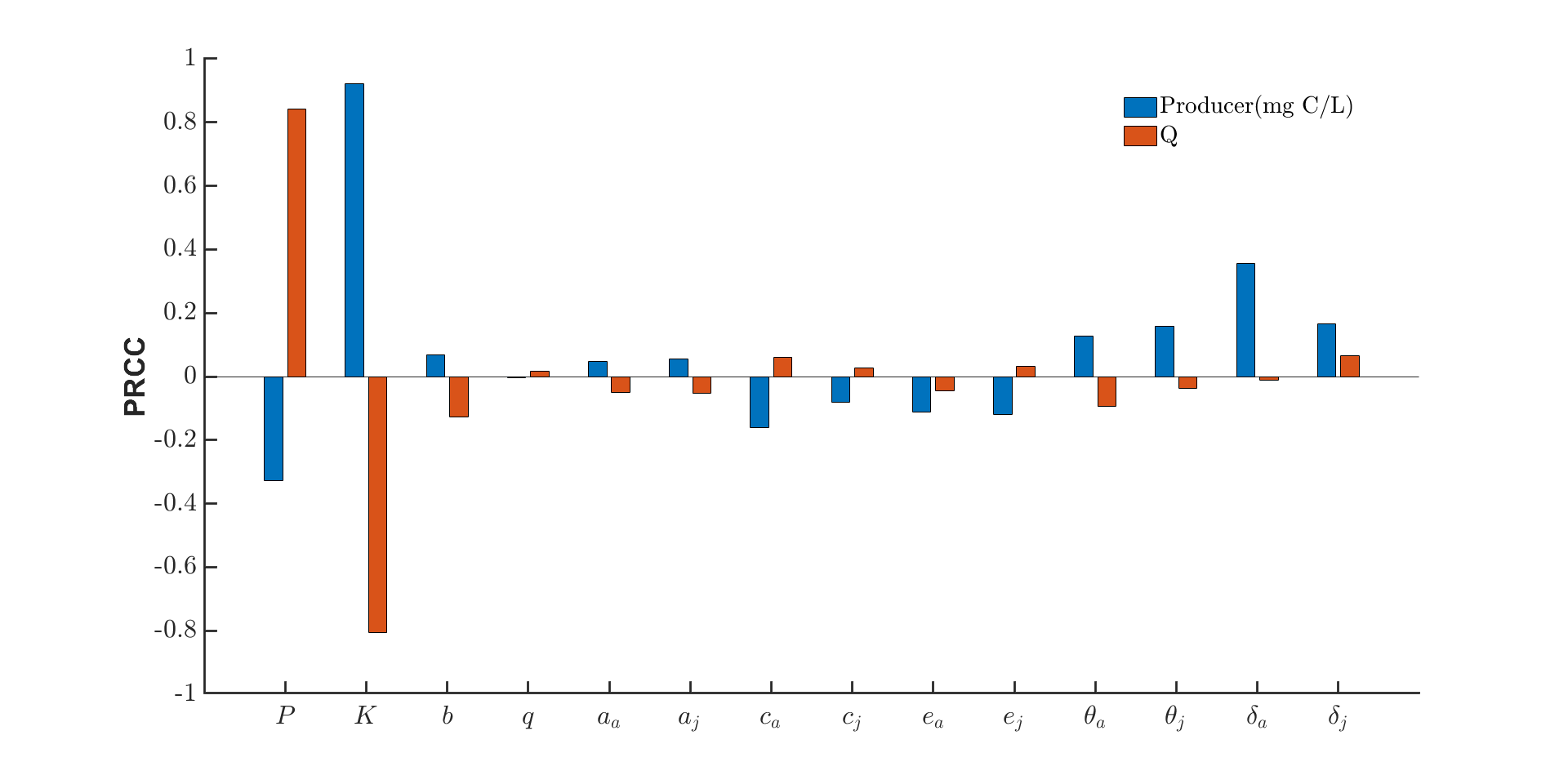}}
\end{minipage}
\hfill
\begin{minipage}{0.49\textwidth}
    \centering
    \subfloat[$c_j$ range 0.3-0.8 (effect on x and Q)]{\includegraphics[width=\textwidth]{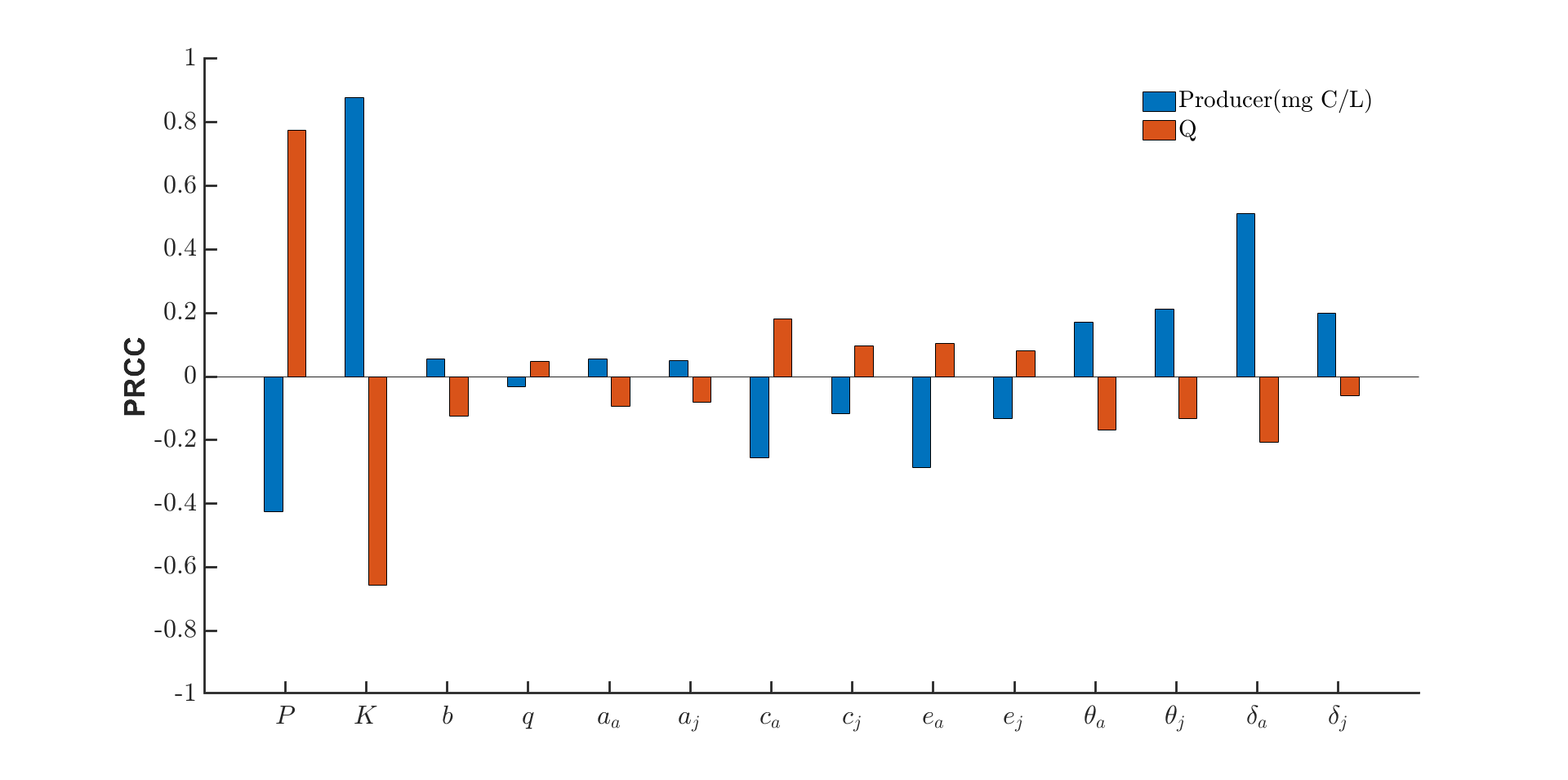}}
\end{minipage}

\vspace{-2pt}

\begin{minipage}{0.49\textwidth}
    \centering
    \subfloat[$\theta_j$ range 0.02-0.04 (effect on x and Q)]{\includegraphics[width=\textwidth]{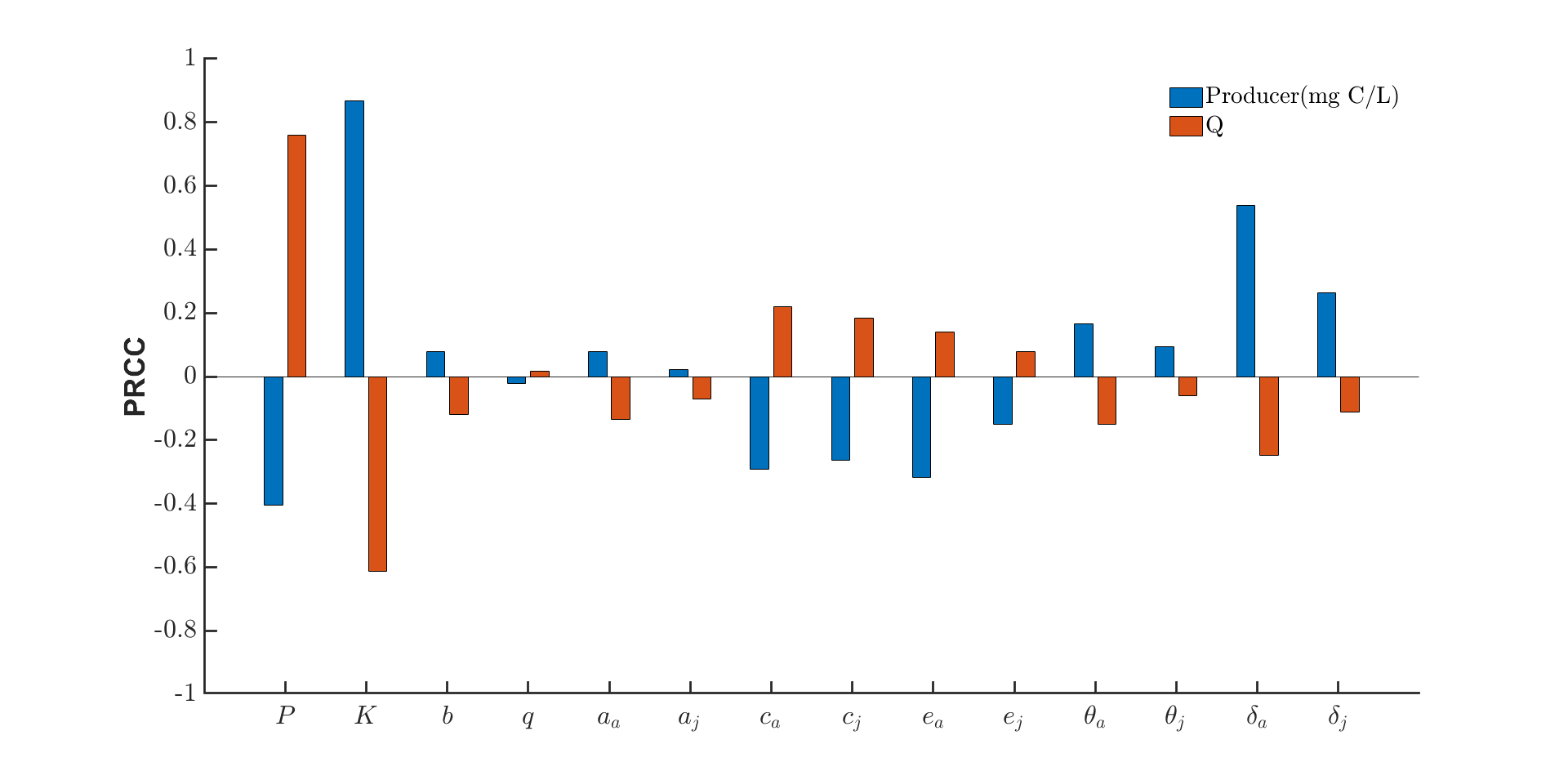}}
\end{minipage}
\hfill
\begin{minipage}{0.49\textwidth}
    \centering
    \subfloat[$\theta_j$ range 0.04-0.08 (effect on x and Q)]{\includegraphics[width=\textwidth]{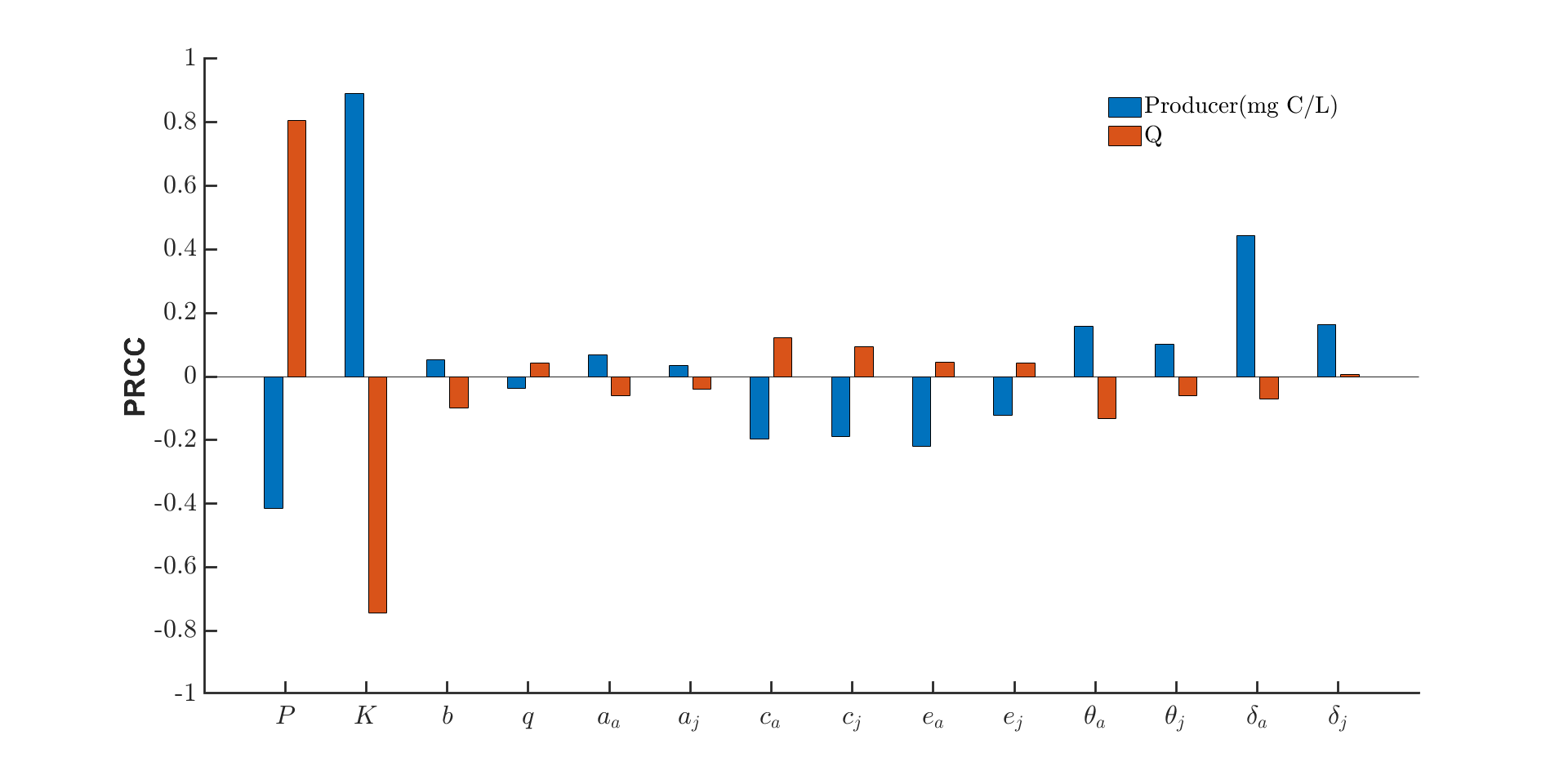}}
\end{minipage}

\vspace{-2pt}

\begin{minipage}{0.49\textwidth}
    \centering
    \subfloat[$\delta_j$ range 0.01-0.24 (effect on x and Q)]{\includegraphics[width=\textwidth]{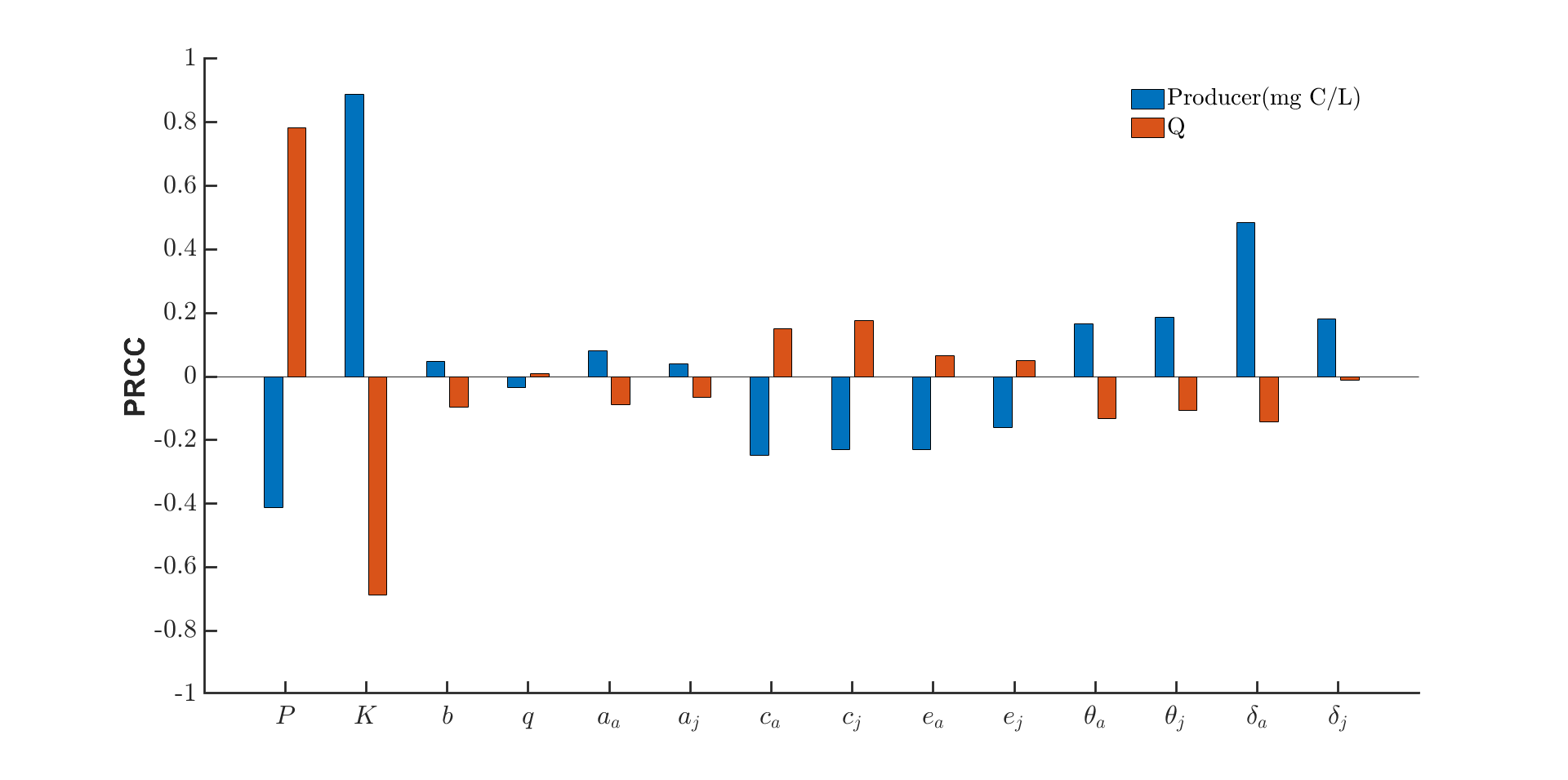}}
\end{minipage}
\hfill
\begin{minipage}{0.49\textwidth}
    \centering
    \subfloat[$\delta_j$ range 0.24-0.3 (effect on x and Q)]{\includegraphics[width=\textwidth]{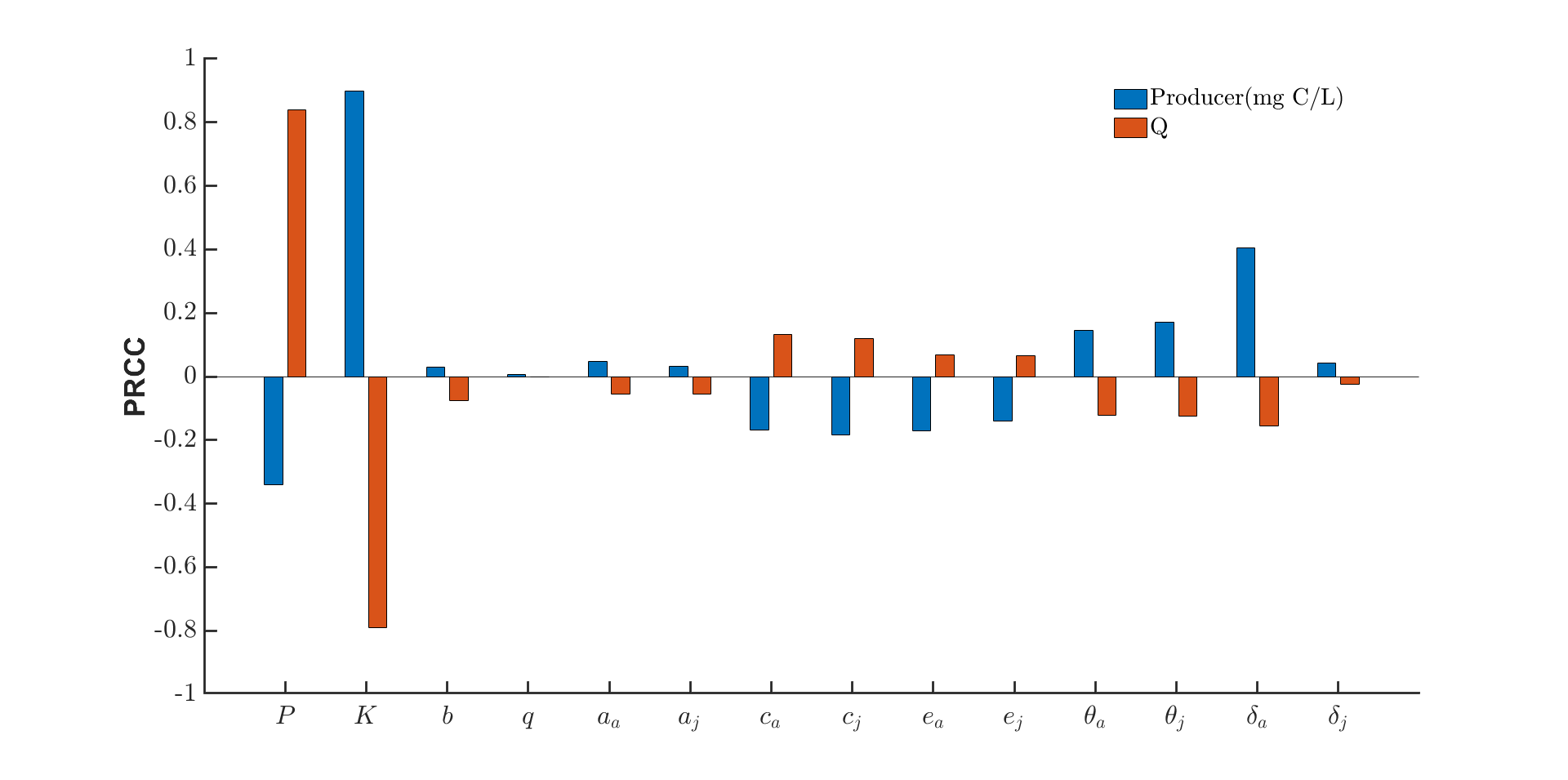}}
\end{minipage}
\vspace{5pt}
\caption{PRCC values of output measures: 
 average producer density, and P:C ratio of producer by using parameter values listed in Table~\ref{tab:params} with 5000 samples. The parameters $c_j$, $\theta_j$, and $\delta_j$ were divided into two distinct monotonic ranges: [0.1  0.3] and [0.3  0.8] for $c_j$, [0.02  0.04] and [0.04  0.08] for $\theta_j$, [0.01  0.24] and [0.24  0.3] for $\delta_j$.}
\label{fig:PRCC_split_x_Q}

\end{figure}

\begin{figure}[H]
\centering

\begin{minipage}{0.49\textwidth}
    \centering
    \subfloat[$c_j$ range 0.1-0.3 (effect on J and A)]{\includegraphics[width=\textwidth]{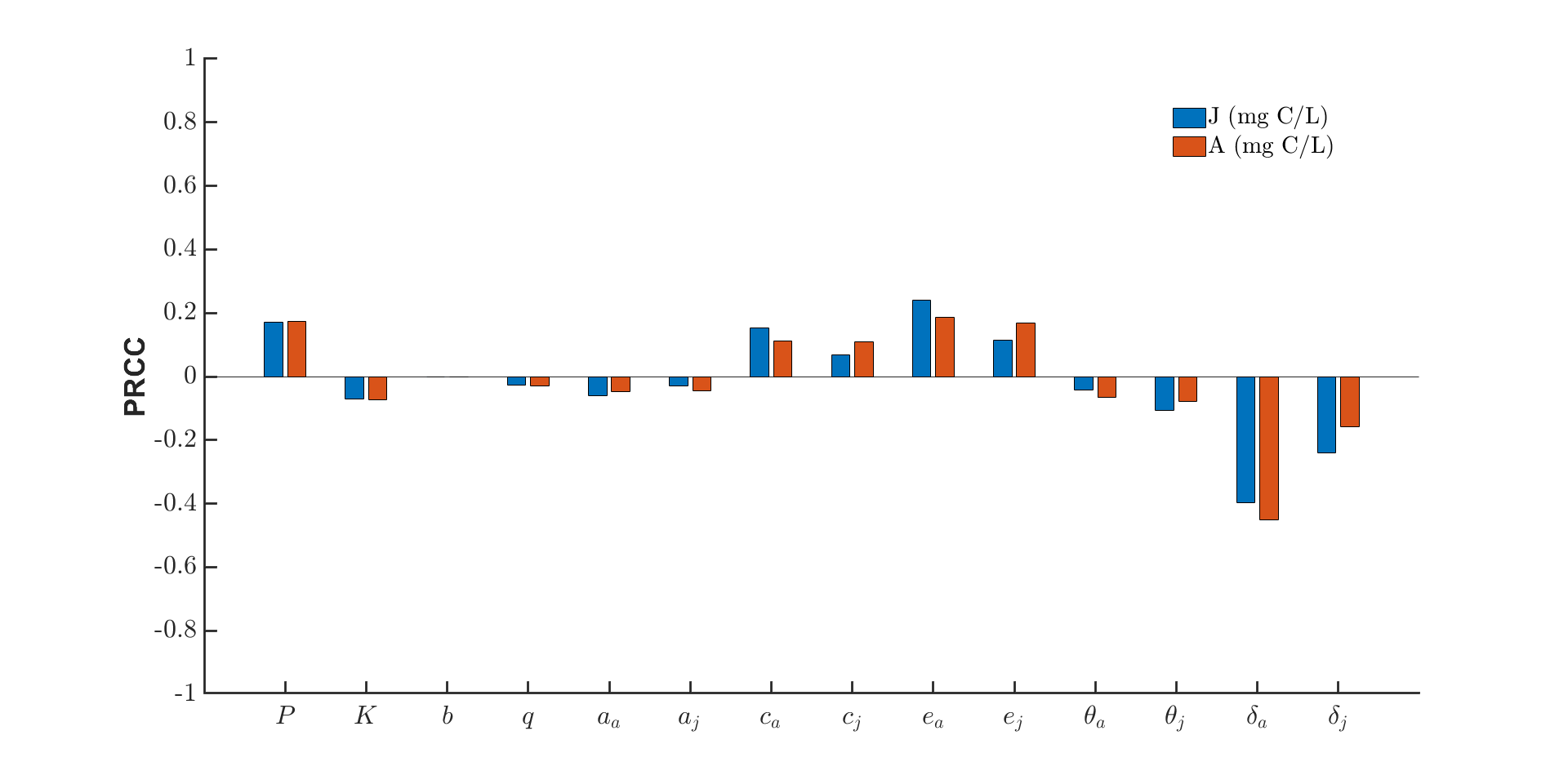}}
\end{minipage}
\hfill
\begin{minipage}{0.49\textwidth}
    \centering
    \subfloat[$c_j$ range 0.3-0.8 (effect on J and A)]{\includegraphics[width=\textwidth]{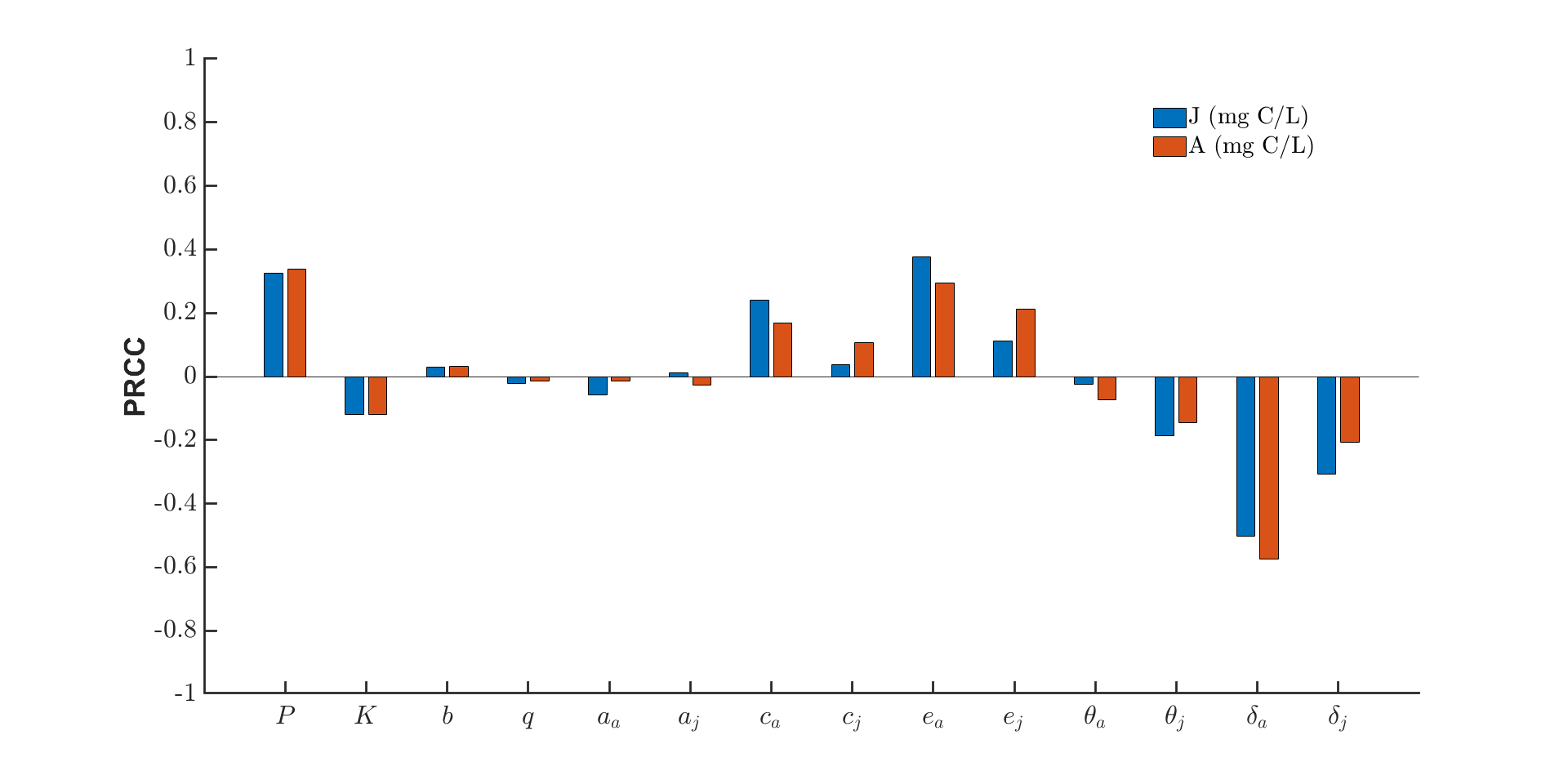}}
\end{minipage}

\vspace{-2pt}

\begin{minipage}{0.49\textwidth}
    \centering
    \subfloat[$\theta_j$ range 0.02-0.04 (effect on J and A)]{\includegraphics[width=\textwidth]{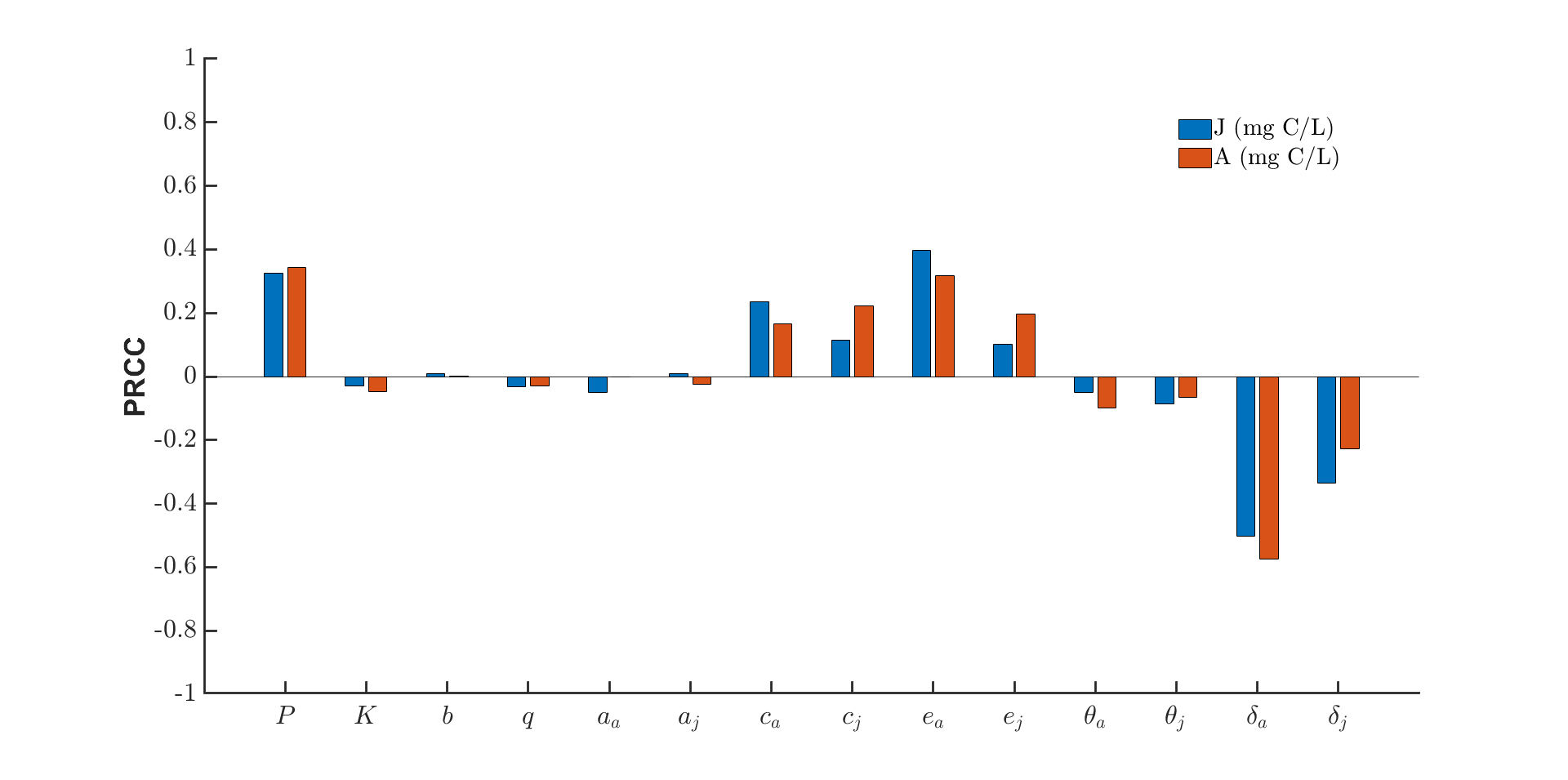}}
\end{minipage}
\hfill
\begin{minipage}{0.49\textwidth}
    \centering
    \subfloat[$\theta_j$ range 0.04-0.08 (effect on J and A)]{\includegraphics[width=\textwidth]{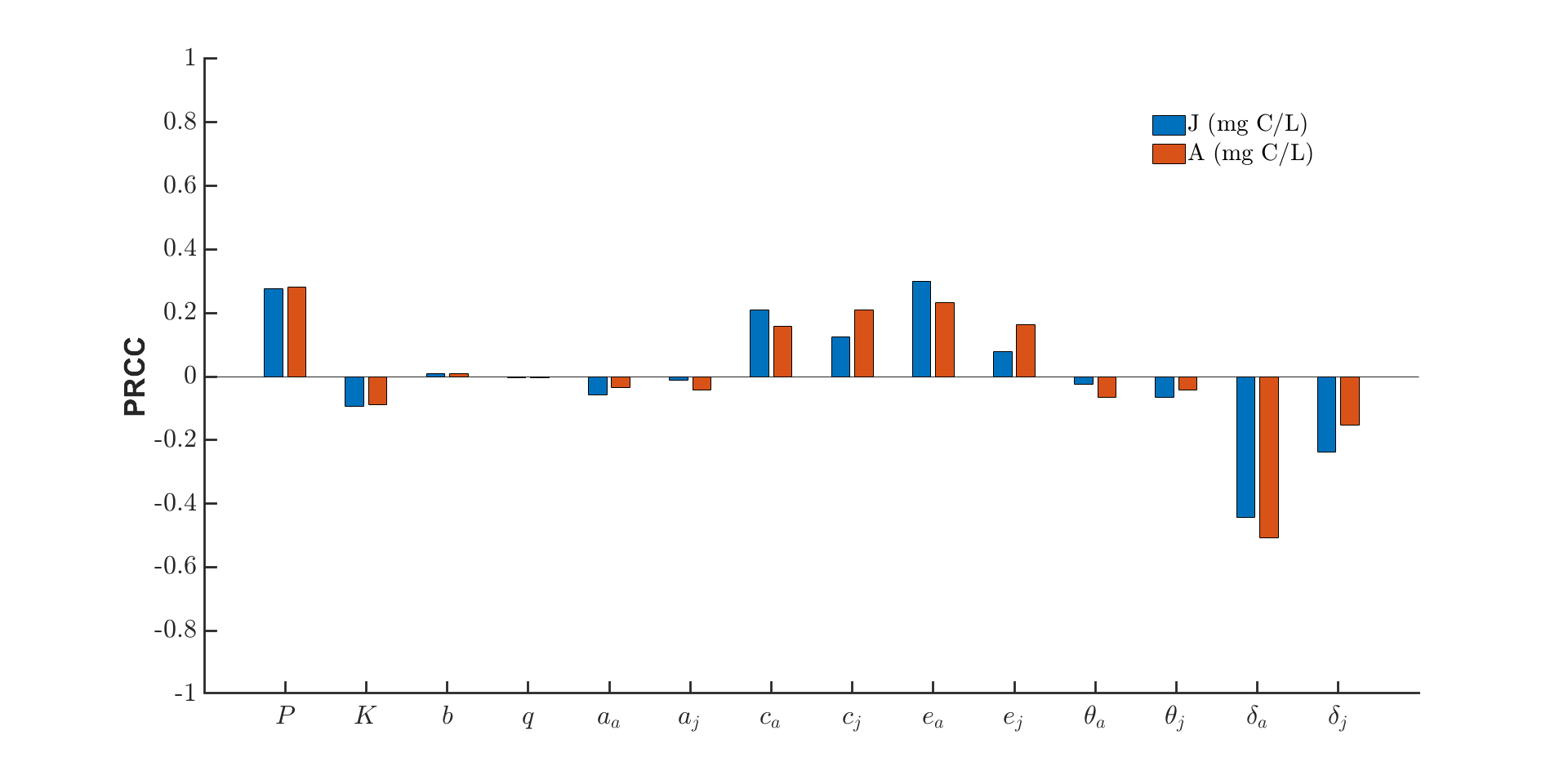}}
\end{minipage}

\vspace{-2pt}

\begin{minipage}{0.49\textwidth}
    \centering
    \subfloat[$\delta_j$ range 0.01-0.24 (effect on J and A)]{\includegraphics[width=\textwidth]{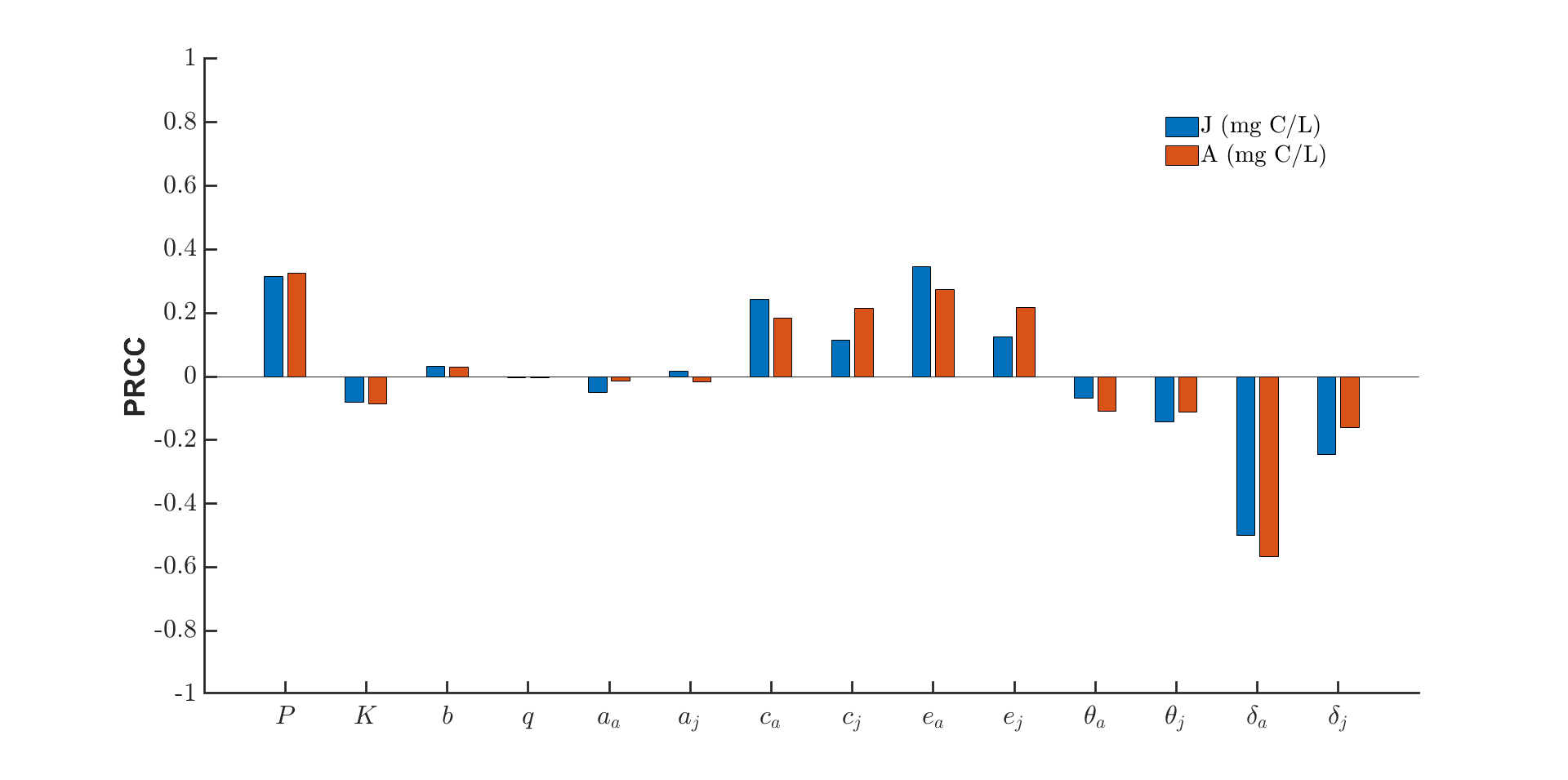}}
\end{minipage}
\hfill
\begin{minipage}{0.49\textwidth}
    \centering
    \subfloat[$\theta_j$ range 0.24-0.3 (effect on J and A)]{\includegraphics[width=\textwidth]{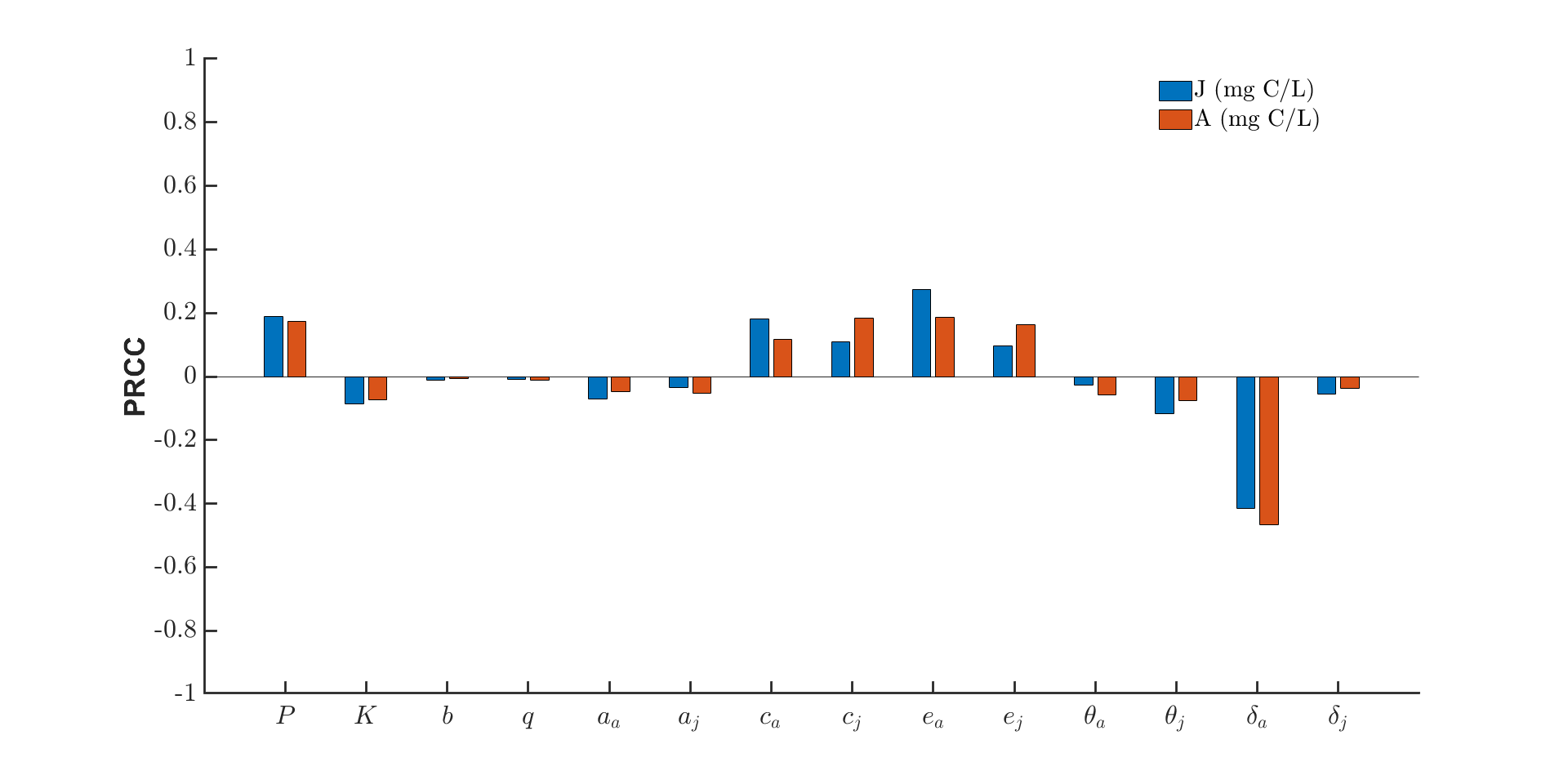}}
\end{minipage}
\vspace{5pt}
\caption{PRCC values of output measures: average juvenile and adult densities by using parameter values listed in Table~\ref{tab:params} with 5000 samples. The parameters $c_j$, $\theta_j$, and $\delta_j$ were divided into two distinct monotonic ranges: [0.1, 0.3] and [0.3, 0.8] for $c_j$; [0.02, 0.04] and [0.04, 0.08] for $\theta_j$; [0.01, 0.24] and [0.24, 0.3] for $\delta_j$. }
\label{fig:PRCC_split_J_A}

\end{figure}

The monotonic relationships between the parameters and output measures are shown in Figs. (\ref{fig:Monotonicity_1}, \ref{fig:Monotonicity_2}). While the majority of the dynamics are monotone there are some non-monotonic trends. However, in some scenarios the variation in the output measure is small enough to justify disregarding monotonicity. For example, the effects of parameter $q$ on output measures: average adult density and $P:C$ ratio of the producer in Fig. \ref{fig:Monotonicity_2}. For other parameters ($P$, $c_a$, $c_j$, $\theta_a$, $\theta_j$, $e_a$, $\delta_a$ and $\delta_j$) the graphs exhibited two distinct monotonic ranges. For each of these cases, we divided these parameter into monotonic ranges, resampled the space, and conducted additional PRCC analyses following techniques from \cite{2008Marino}. 
This new PRCC analysis yielded similar results. For example, for the parameters $c_j$, $\theta_j$, and $\delta_j$,  the graphs were divided into the following monotonic ranges: [0.1  0.3] and [0.3  0.8] for $c_j$,  [0.02  0.04] and [0.04  0.08] for $\theta_j$,  [0.01  0.24] and [0.24  0.3] for $\delta_j$, and these PRCC analyses (Figs. \ref{fig:PRCC_split_x_Q},\ref{fig:PRCC_split_J_A}) align with the original sensitivity analyses on the full parameter ranges (Figure \ref{fig:PRCC}).



\setlength{\bibsep}{0pt}
\bibliographystyle{abbrv} 
\bibliography{Sources}

\end{document}